\documentclass[12pt,preprint]{aastex}

\usepackage{amsmath}
\begin{document}
\title{On the orbital evolution of a giant planet pair embedded in a gaseous disk. I: Jupiter-Saturn configuration}

\author{Hui Zhang  \& Ji-Lin Zhou }
\altaffiltext{}{Department of Astronomy \& Key Laboratory of Modern Astronomy
and Astrophysics in Ministry of Education£¬Nanjing University, Nanjing
210093,China ; huizhang@nju.edu.cn}

\begin{abstract}
We carry out a series of high resolution ($1024\times 1024$) hydrodynamical
simulations to investigate the orbital evolution of Jupiter and Saturn embedded
in a gaseous protostellar disk. Our work extends the results in the classical
papers of Masset \& Snellgrove (2001) and Morbidelli \& Crida (2007) by
exploring various surface density profiles ($\sigma$), where $\sigma \propto
r^{-\alpha}$. The stability of the mean motion resonances(MMRs) caused by the
convergent migration of the two planets is studied as well. Our results show
that:(1) The gap formation process of Saturn is greatly delayed by the tidal
perturbation of Jupiter. These perturbations cause inward or outward runaway
migration of Saturn, depending on the density profiles on the disk. (2) The
convergent migration rate increases as $\alpha$ increases and the type of MMRs
depends on $\alpha$ as well. When $0<\alpha<1$, the convergent migration speed
of Jupiter and Saturn is relatively slow, thus they are trapped into 2:1 MMR.
When $\alpha>4/3$, Saturn passes through the $2:1$ MMR with Jupiter and is
captured into the $3:2$ MMR. (3) The $3:2$ MMR turns out to be unstable when
the eccentricity of Saturn ($e_s$) increases too high. The critical value above
which instability will set in is $e_s \sim 0.15$. We also observe that the two
planets are trapped into $2:1$ MMR after the break of $3:2$ MMR. This process
may provide useful information for the formation of orbital configuration
between Jupiter and Saturn in the Solar System.
\end{abstract}

\section{Introduction}
Planet-planet interaction within an environment of gas disk is an important
procedure that may account for the initial conditions of multiple planet system
after the depletion of gas disk, and thus affects their final orbital
configurations. One of the notable configuration that two planets may achieve
is the mean motion resonance (MMR). For example, the crossing of 2:1 MMR
between Jupiter and Saturn was proposed to account for the later heavy
bombardments of the Solar system\citep{Tsi05,Gom05}. MMRs are also common in
the recently detected exoplanets{\footnotetext{http://exoplanet.eu/}}. Among
the $\sim 45$ detected multiple exoplanet systems, at least 7 planet pairs are
believed to be trapped in low order MMRs (see \textbf{Table} \ref{table 1} for
a list).

According to the general theory of disk-planet interaction, a single planet
embedded in a gaseous disk may undergo various types of migration. For a planet
with mass smaller than several Earth masses ($M_\oplus$), the angular momentum
exchange between it and the gas disk will cause a net momentum loss on the
planet and results in a fast orbit decay, which is called type I
migration\citep{Gol79,War97}. For a planet with mass comparable to Jupiter, it
opens a gap around its orbit. Through the planet, the angular momentum exchange
between the outer and inner part of the disk balances each other. This effect
locks the planet and forces it to move as a part of the disk, at the viscous
evolution timescale. This is called the type II migration \citep{Lin86a}.
Moderate mass planets(comparable to Saturn) may undergo very fast migration,
which is believed to be caused by the large corotation torque risen from the
perturbed coorbital zone of the planet. It is now named runaway migration or
type III migration with timescale of several tens of orbit periods
only\citep{Mas03}.

Due to the different migration rates of two planets embedded together in a
disk, MMRs may be established through the relative convergent migration. For
example, when the outer planet is less massive than the inner one and undergoes
type I or III migration, it will probably catch up the more massive inner one
which undergoes type II migration, even they are both migrating at the same
direction. So, in multiple planet systems with heavier planet locating at the
inner side, planets may be easily trapped into the MMRs. Many constructive
studies have been made to investigate this scenario, either by three-body
simulations with prescribed disk effects, e.g., \citet{Sne01} and
\citet{Nel02}, or through hydrodynamic simulations, e.g., \citet{Pap05},
\citet{Kle04,Kle05} and \citet{Pie08}.

\citet*{Mas03} investigated a case where the Jupiter(inner)-Saturn(outer) pair
embedded in a protostellar disk. In that case, Saturn migrates inward very
fast(type III migration) and is then captured into $3:2$ MMR by Jupiter which
undergoes slow type II migration. As soon as the resonance is well established,
the two planets reverse their migration and move outward together, preserving
their resonance. \cite*{Mor07} confirmed these results by using a more reliable
code that describes the global viscous evolution of the disk. They considered a
wide set of initial conditions and found the $3:2$ MMR is a robust outcome of
Jupiter-Saturn pair, which is compatible to the requirement of a compact
initial configuration of Jupiter and Saturn (with period ratio slightly less
than 2). After gas depletion, the subsequent 2:1 MMR crossing under the
interaction with planetesimal disk may achieve the present configuration of
Solar system\citep{Tsi05,Gom05,Mor05}.

\cite*{Mor07} also showed that, the common migration rate of Jupiter and Saturn
depends on the viscosity of the disk, as well as vertical scale height ($H/r$)
of the gas disk.  At $H/r=0.05-0.06$, Jupiter and Saturn seem to be in a
quasi-stationary configuration without migration. With such conditions,
\citet{Mor07b} found that, the Jupiter-Saturn pair could maintain the $3:2$ MMR
for at least $1500$ Jovian orbital periods when the gas disk dissipating slowly
and smoothly. And we note that, through their evolutions, the eccentricities of
Jupiter and Saturn stay at relatively low values($e<0.05$).

In this paper, following the work of \citet{Mas01} and \citet{Mor07}, we
investigate the orbital evolution of Jupiter-Saturn pair embedded in gas disk
with different slope of surface density ($\alpha$), i.e., $\alpha=-\ln
(\sigma)/\ln (r)$.  We will show that, under suitable disk condition, $\alpha
\ge 4/3 $ (or $<1$), Jupiter and Saturn will undergo outward (or inward)
migration after trapped into $3:2$ (or $2:1$, respectively) MMRs. Thus the
surface density profile plays an important role on determining the type of
resonance and their consequent migrations. Also we will show that, under some
circumstances, the eccentricities of Jupiter and Saturn will be excited up to
0.15 along the migration. The eccentricity excitation and subsequent stability
of the MMRs between Jupiter and Saturn are discussed. Such a study help us to
reveal the orbital architecture formation of our Solar system.

Our paper is organized as follows: in Section 2 we describe our model
including the numerical methods and the initial settings. In Section 3 we
present our results as well as the analysis. The conclusions and discussions
will be placed at Section 4.

\section{Model and Numerical Set up}
\subsection{Physical model}
Following conventional procedures, we simulate the full dynamical interaction
of a system includes a solar type star, two giant protoplanets and a
2-dimensional (2D) gas disk. The star is fixed at the origin of the system with
both the planets and disk surrounding it. We denote the inner and outer planets
with subscript 1 and  2, respectively.

For numerical convenience, the gravitational constant $G$  is set to be $1$.
The Solar mass ($M_{\odot}$) and the initial semi-major axis of inner planet
($R_{10}=5.2$ AU) are set as the units of mass and length, respectively. Then
the unit of time is $\frac{1}{2\pi}P_{10}$, where $P_{10}$ the initial period
of inner orbit. The mass of central star is set to be 1 Solar mass ($M_*=1
M_\odot$).

We solve the continuity and momentum equations of the gas disk in 2-D space,
neglecting the self-gravity of the gas. In order to properly describe the
global viscous evolution of the whole disk and to avoid the annoying inner
boundary, we employ a Cartesian grid. The vertically averaged continuity
equation for the gas is given by
\begin{equation}
\frac{\partial \sigma}{\partial t} + \frac{\partial(\sigma u_{x})}{\partial x} + \frac{\partial(\sigma
u_{y})}{\partial y} = 0,
\end{equation}
where $\sigma$ is the surface density.
The equations of motion in the Cartesian coordinates are
\begin{equation}
\frac{\partial(\sigma u_{x})}{\partial t}+\frac{\partial(\sigma
u_{x}^{2})}{\partial x}+\frac{\partial(\sigma u_{x} u_{y})}{\partial y} =
-\frac{\partial P}{\partial x}-\sigma \frac{\partial \Phi}{\partial x}+
\frac{\partial Q_{xx}}{\partial x}+\frac{\partial Q_{xy}}{\partial y},
\end{equation}

\begin{equation}
\frac{\partial(\sigma u_{y})}{\partial t}+\frac{\partial(\sigma u_{x}
u_{y})}{\partial x}+\frac{\partial(\sigma u_{y}^{2})}{\partial y} =
-\frac{\partial P}{\partial y}-\sigma \frac{\partial \Phi}{\partial y}+
\frac{\partial Q_{xy}}{\partial x}+\frac{\partial Q_{yy}}{\partial y},
\end{equation}
where $P$ is the pressure and $\Phi$ is the gravity potential of the whole
system. $\Phi$ includes the softened potential of the central star $\Phi_{*}$,
softened potential of the giant planets ($\Phi_{p,1}$,$\Phi_{p,2}$) and the
indirect potential $\Phi_{\rm ind}$ rises from the acceleration of the origin,
which is caused by the planets and the gas disk. We adopt a softened gravity
potential of the central star to avoid the singularity at the very center of
the computational domain,
\begin{equation}
\Phi_{s}=-\frac{GM_\odot}{\sqrt{x^2+y^2+\epsilon_{*}^2}}.
\end{equation}
The potential of each planet is also softened,
\begin{equation}
\Phi_{p,i}=-\frac{GM_{p,i}}{\sqrt{(x-x_{p,i})^2+(y-y_{p,i})^2+\epsilon_{p}^2}},~~(i=1,2),
\end{equation}
where $\epsilon_{*}$ and $\epsilon_{p}$ are the soften length to the central star
and planets respectively. We test several values of $\epsilon_{*}$ and set
it to be $0.05$ for the balance of efficiency and accuracy. The $\epsilon_{p}$
is set to be $0.6H$, where $H$ is the scale height of the gas disk.

The viscosity of gas is carried out by solving the stress tensors $Q_{xx}$,
$Q_{yy}$ and $Q_{xy}$ explicitly:
\begin{align}
Q_{xx}=2\eta[\frac{\partial u_x}{\partial x}-\frac{1}{3}(\frac{\partial
u_x}{\partial x}+\frac{\partial u_y}{\partial y})],\\
Q_{yy}=2\eta[\frac{\partial u_y}{\partial y}-\frac{1}{3}(\frac{\partial
u_x}{\partial x}+\frac{\partial u_y}{\partial y})],\\
Q_{xy}=\eta(\frac{\partial u_x}{\partial y}+\frac{\partial u_y}{\partial x}).
\end{align}
where $\eta=\sigma\nu$ and $\nu$ is the dynamical viscous coefficient of the
gas, which is assumed to be constant all over the disk.

We assume the disk gas has a polytropic equation of state:
\begin{equation}
P=K \sigma^\gamma,
\end{equation}
where $\gamma=\frac{5}{3}$. $K$ is a constant that makes the following equation
satisfied at $r=1$,
\begin{equation}
c_s(r)=\sqrt{\frac{\partial P}{\partial \sigma}}=\left(\frac{H}{r}\right) v_{\rm kep}|_{r=1},
\end{equation}
where $c_s(r)$ is the speed of sound and $v_{\rm kep}(r)$ is the Keplerian velocity
of the gas. According to our settings and units, $K=0.1349$. To
focus on the effects of the surface density distribution, we fix the
disk viscosity  $\nu=10^{-6}$ and the disk aspect ratio $H/r=0.04$.
The planetary accretion and self-gravity of gas disk are neglected to reduce
variables and to improve the computational efficiency.

\subsection{Mesh configuration and computational domain}
\begin{figure}
\epsscale{0.6} \plotone{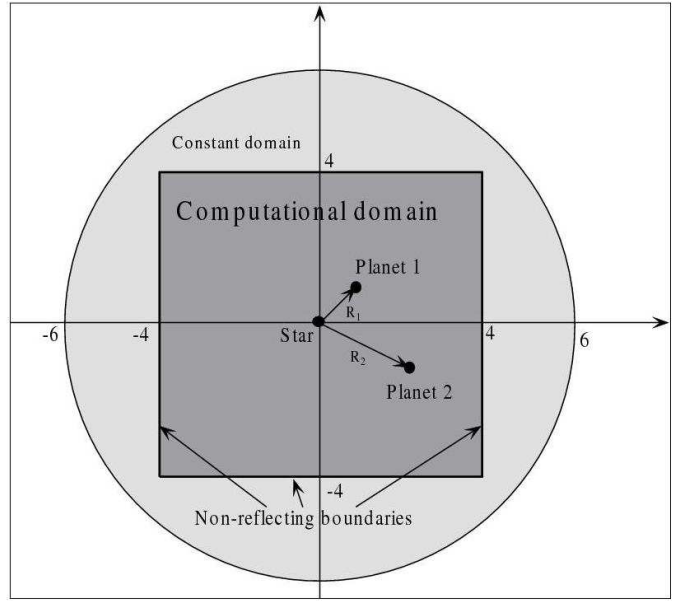} \caption{The computational domain
is from -4 to 4 in x direction and from -4 to 4 in y direction(gray square).
Surrounding it are four non-reflecting boundaries. Area outside the square is
assumed to stay constant. We take the gravity comes from the whole round
area($R\leq 6$) as a background potential which is a function of radius.
\label{figure 1}}
\end{figure}

The \emph{Antares} code we have developed is adopted in the simulations. It is
a 2D Godunov code based on the exact Riemann solution for isothermal or
polytropic gas, featured with non-reflecting boundary conditions. The detail of
this method was described elsewhere\citep{Yua05}. Since we adopt the
Cartesian grids, the computational domain is also a square. \textbf{Figure}
\ref{figure 1} shows the computational domain. To get
the proper gravity potential of a disk, we add an circular area outside the
computational domain. Since  it is far from the interested place,
 this circular area stays at the initial condition during the simulations and its gravity
potential is pre-calculated. The computational domain (gray area) is divided by
a Cartesian mesh. The resolution of this mesh is a crucial issue for
hydrodynamics simulations, since poor resolution may introduce non-physical
effects that affects the migration of planets greatly. So we employ a high
resolution: $N_x\times N_y=1024\times 1024$. The real computational time for a
single 2000-orbits run is $3-4$ weeks,   in a full parallelized 64-cpus
cluster. This almost reaches the upper limit of normal computational ability,
and the higher resolution may be achieved by the nested grids or other
technics.

The reasons and advantages of our choosing of Cartesian grids had been
discussed in our previous paper\citep{Zha08}. For the orbits of two planets we
integrate a three-body problem associate with the potential of gas disk by
adopting a 8th-order Runge-Kutta  integrator. The global time step is set as
the minimum of the hydrodynamical  and the orbital integration part.

\subsection{Corner-Transport Correction}
\begin{figure}
\epsscale{0.6} \plotone{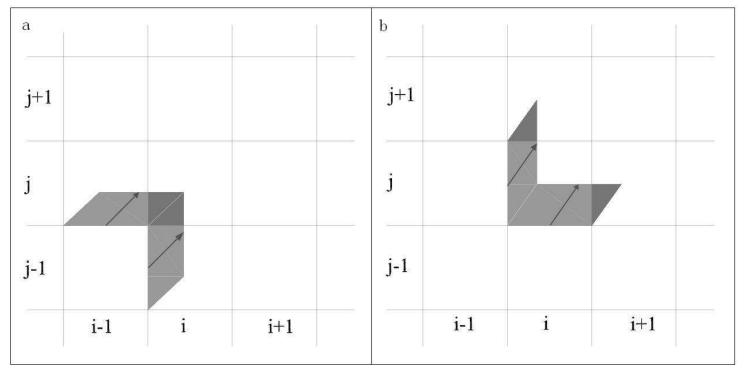} \caption{The real flux flows at an
angle to the grid lines(assume $u_x>0,u_y>0$). (a) The new value
$Q^{n+1}_{i,j}$ in cell $(i,j)$ should also be affected by the old value
$Q^{n}_{i-1,j-1}$ in cell $(i-1,j-1)$. (b) The flux at four interfaces of cell
$(i,j)$ need additional corrections---see the four dark triangles.
\label{figure 2}}
\end{figure}

Although it is easy to implement, the Cartesian grids have a disadvantage to
simulate a circularly orbiting gas disk, where the physical stream lines are
 neither parallel nor perpendicular to either of the grid lines in x-
and y-directions. The flux is evaluated at each interface at which the velocity
of gas projects to the coordinate directions. When the grid resolution is low,
the conservative law may break along the grid lines(see \textbf{Figure}
\ref{figure 2}) and much dissipation arise. It is a kind of numerical viscosity
which reaches maximum at the diagonal area of the computational domain. To make
this numerical viscosity negligible, the resolution of the grids usually needs
to be very high. Instead, we adopt the CTU(Corner-Transport Upwind) method to
minimize this grid effect at the resolution that we can achieve so far.

Following the CTU method, we add correcting terms to the flux at each interface
of all cells. The value of this correcting flux depends on the angle between
the direction of the physical velocity and the grid lines. As \textbf{Figure}
\ref{figure 2} shows, by assuming $u_x>0$ and $u_y>0$, the cell $(i,j)$ is
affected by an additional flux comes from cell $(i-1,j-1)$. The cell averaged
value $Q^{n+1}_{i,j}$, for example, is modified by a term,
\begin{equation}
\frac{\frac{1}{2}u_x u_y(\Delta t)^2}{\Delta x\Delta
y}(Q^{n}_{i-1,j}-Q^{n}_{i-1,j-1}),
\end{equation}
where $\frac{1}{2}u_x u_y (\Delta t)^2$
is the area of the small triangular portion moving into cell $(i,j+1)$ and
$\Delta x\Delta y$ is the area of the cell in which the jump
$Q^{n}_{i,j}-Q^{n}_{i-1,j}$ is averaged. The corresponding fluxes evaluated at
the four interfaces of cell $(i,j)$ now read
\begin{flalign}
\begin{split}
F_{i-\frac{1}{2},j}=F_{i-\frac{1}{2},j}-\frac{1}{2}\frac{\Delta
t}{\Delta y}u_x u_y(Q^n_{i-1,j}-Q^n_{i-1,j-1}),\\
F_{i+\frac{1}{2},j}=F_{i+\frac{1}{2},j}-\frac{1}{2}\frac{\Delta
t}{\Delta y}u_x u_y(Q^n_{i,j}-Q^n_{i,j-1}),\\
G_{i,j-\frac{1}{2}}=G_{i,j-\frac{1}{2}}-\frac{1}{2}\frac{\Delta
t}{\Delta y}u_x u_y(Q^n_{i,j-1}-Q^n_{i-1,j-1}),\\
G_{i,j+\frac{1}{2}}=G_{i,j+\frac{1}{2}}-\frac{1}{2}\frac{\Delta t}{\Delta y}u_x u_y(Q^n_{i,j}-Q^n_{i-1,j}).
\end{split}
\end{flalign}
The details and implementations of CTU method for conservation laws have been
well discussed by \citet{Col90}.

\subsection{Comparison with other codes}
To ensure the reliability of our code, we performed a series of comparisons
with other representative codes, e.g. FARGO. We examined the gap opening by
Jupiter in a 2D gas disk. The physical setups and initial conditions are the
same with that adopted by \citet{Val06}, except that our radial domain is from
$0$ to $2.5$ instead of $[0.4, 2.5]$. The grid resolution is $640\times640$ in
the spatial range of
$[x_{min},x_{max}]\times[y_{xmin},y_{max}]=[-2.5,2.5]\times[-2.5,2.5]$.
Following their descriptions, we focus on the density contours of the gap, the
evolution of density profiles, the evolution of total mass and the torques
exerted on Jupiter. All the comparisons are performed in both inviscid and
viscous disk, where the dynamical viscous coefficient is set to be $\nu=0$ and
$\nu=10^{-5}$ respectively. The results with which we compared ourselves are
obtained from the
web\footnotemark{\footnotetext[1]{http://www.astro.su.se/groups/planets/comparison/}}
which maintained by de Val-Borro.

\begin{figure}
\epsscale{0.6} \plotone{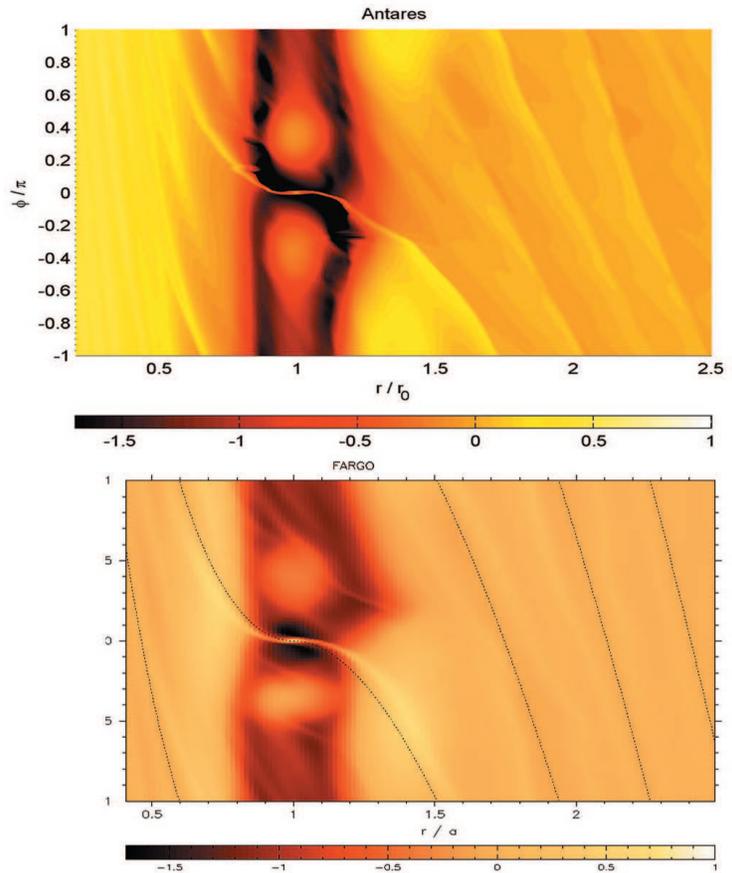} \caption{ Density maps in
logarithmic scale after 100 orbits for the inviscid simulations. Note that the
computational domain of \emph{Antares} is $(r_{min},r_{max})=(0,2.5)$, instead
of $(0.4,2.5)$. The density range is $-1.7 <\log(\sigma/\sigma_0)< 1$.
\emph{Antares} adopts isothermal equation of state(EOS) instead of the locally
isothermal EOS, so the pitch angle is bit large than that in FARGO. The results
of FARGO are obtained from the web:
http://www.astro.su.se/groups/planets/comparison, which is maintained by de
Val-Borro. \label{figure 17}}
\end{figure}

\begin{figure}
\epsscale{0.6} \plotone{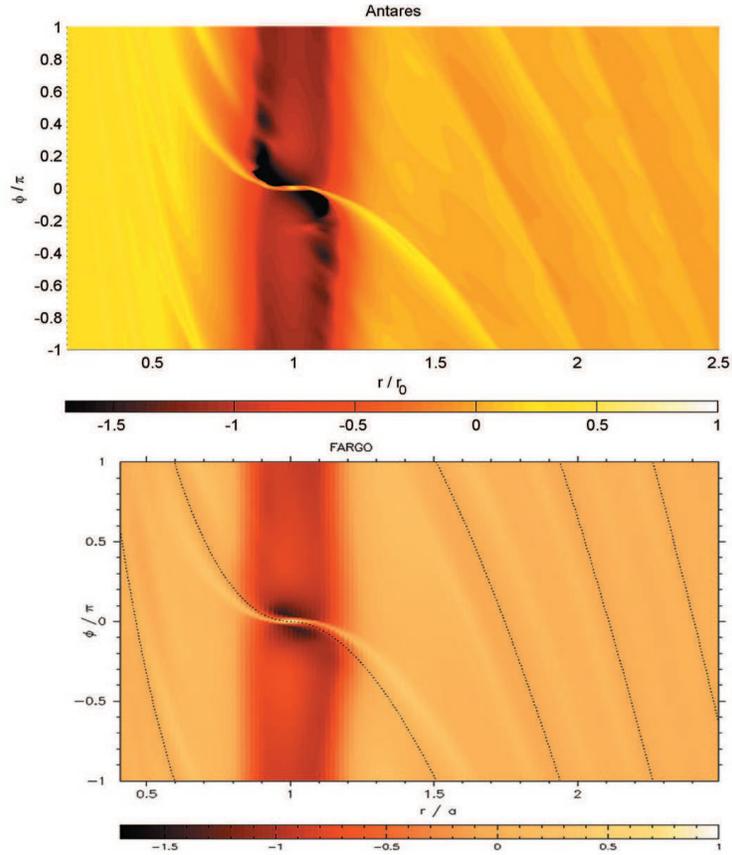} \caption{ Density maps in
logarithmic scale after 100 orbits for the viscous simulations. The viscous
coefficient $\nu=10^{-5}$. The density range is also $-1.7
<\log(\sigma/\sigma_0)< 1$. The results of FARGO are obtained from the web:
http://www.astro.su.se/groups/planets/comparison, which is maintained by de
Val-Borro. \label{figure 18}}
\end{figure}

\textbf{Figure} \ref{figure 17} shows the density contours after 100 orbits for
the inviscid simulation. Two shocks are observed in our simulation
(\emph{Antares}): the primary one starts from the planet's location and the
secondary one starts near the $L_5$ point. We find the pattern of the spiral
arms are similar with that of the other codes although the pitch angle of the
primary arm(outside the orbit of planet) is a bit higher in our simulation,
which is probably caused by the relative large sound speed their. Since the
exact Riemann solution is not valid for locally isothermal gas, we adopt the
full isothermal equation of state in this comparison tests instead and set the
sound speed $c_s=\frac{H}{r} v_{kep}|_{r=1}$, where $H$ is the height of disk
and $v_{kep}$ is the Keplerian orbit speed. In the gap, there are two symmetric
density enhancements locate close to the $L_4$ and $L_5$ points at azimuthal
distance $\Delta\phi=\pm\pi/3$ from the planet's location. This is in good
agreement with the theoretical prediction and the results of FARGO. We also
notice there is a density bump, which indicates the congregation of vortensity,
orbiting along the outer edge of the gap, which is observed in many other codes
as well, e.g. the results of FARGO and NIRVANA-GDA\citep{Val06}. In the viscous
simulation, the gap opened by the planet is narrower and
smoother(\textbf{Figure} \ref{figure 18}). The density enhancements seen at the
Lagrangian points inside the gap in the inviscid simulation are dissipated, so
is the density bump at the outer edge of the gap. Our simulation presents the
same results with the other codes and shows more detailed structures within the
gap.

\begin{figure}
\epsscale{0.6} \plotone{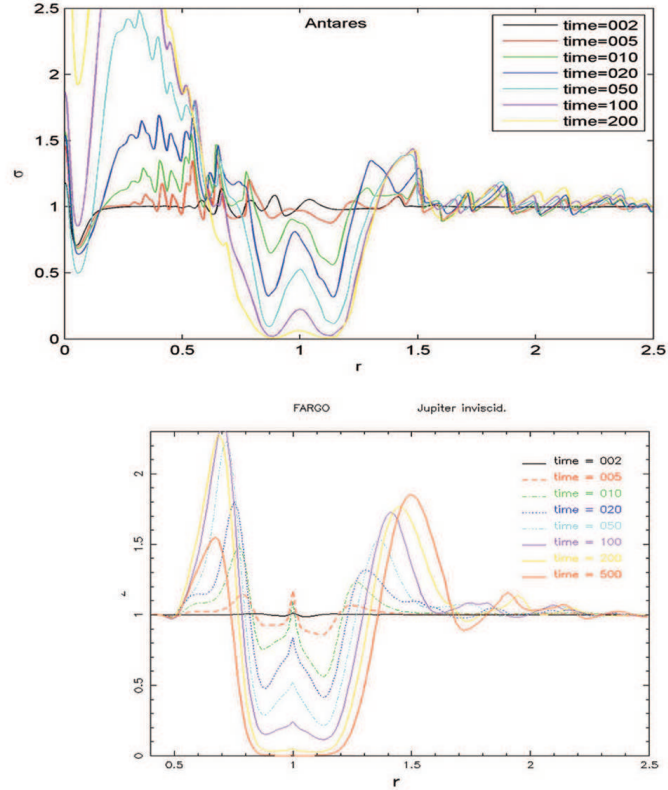} \caption{The normalized
surface density profiles averaged azimuthally over $2\pi$ after 100 orbits for
the inviscid simulations. Note that the computational domain of \emph{Antares}
is $(r_{min},r_{max})=(0,2.5)$, instead of $(0.4,2.5)$. The surface density at
inner disk increases and keeps a high level when the inner open boundary is
absent. The results of FARGO are obtained from the web:
http://www.astro.su.se/groups/planets/comparison, which is maintained by de
Val-Borro. \label{figure 19}}
\end{figure}

\begin{figure}
\epsscale{0.6} \plotone{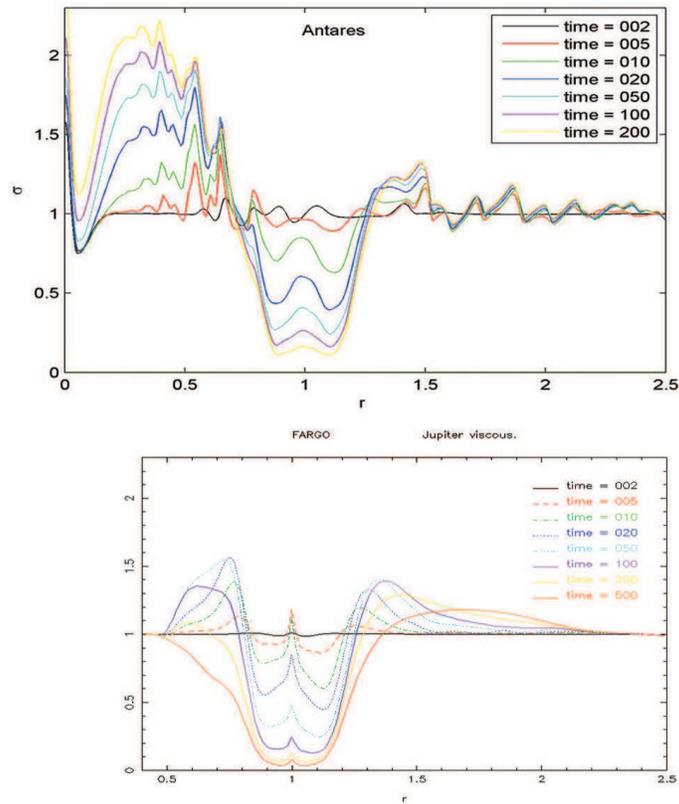} \caption{ The normalized
surface density profiles averaged azimuthally over $2\pi$ after 100 orbits for
the viscous simulations. The viscous coefficient $\nu=10^{-5}$. \emph{Antares}
shows the proper behavior of gas under the increasing dissipation effect---the
gap becomes narrower and shallower. The results of FARGO are obtained from the
web: http://www.astro.su.se/groups/planets/comparison, which is maintained by
de Val-Borro. \label{figure 20}}
\end{figure}

\textbf{Figure} \ref{figure 19} and \ref{figure 20} show the density profiles
at different times in the inviscid and viscous simulations respectively. The
width and depth of the gap in our simulations are in good agreement with those
in FARGO. Our code also present the proper diffusion effects---decreasing on
both the width and depth of the gap in viscous simulation. The major difference
is the density on the inner disk: the surface density on the inner disk
increases to higher value in our simulation than that in FARGO. This mainly
comes from the different treatments at the center of disk.

As the Cartesian grids are adopted, we need \emph{not} introduce any inner
boundary at the center of disk but just let the gas accumulate there. The
competition between the increasing pressure, the dissipation of gas and the
gravity of central star(there also could be the accretion of the central star
which is not included in this comparison simulation) will result in an
equilibrium and maintain an inner structure naturally(whose scale is around
$r\leq0.1$ and changes with time). In a real proto-stellar disk, there probably
exists an inner boundary near the center of disk, however it should be
maintained by some equilibriums, e.g. the evaporation and the refilling of gas,
and should move inward or outward according to the changes of local situation,
e.g. the enhancement of the density, instead of a fixed boundary. And further
more, a full inner disk(includes the part $r<0.4$) plays a great role on the
dynamical interaction between the giant planet and the global disk, and thus
can not be ignored without careful treatments\citep{Cri08}.

\begin{figure}
\epsscale{0.6} \plotone{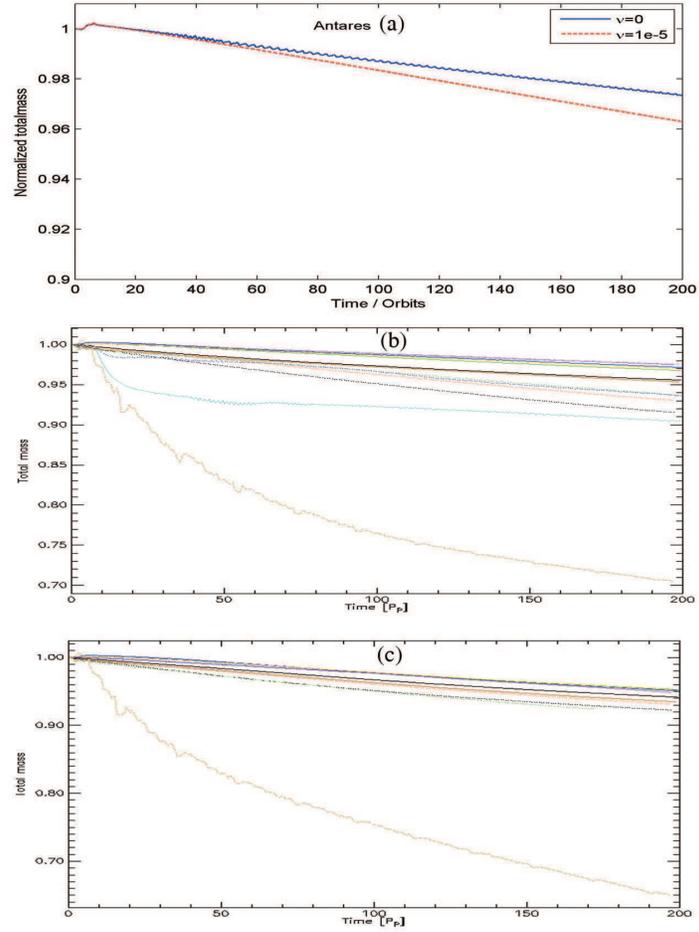} \caption{The evolution of total
mass contained in the disk. (a) Total mass of disk calculated by \emph{Antares}
in inviscid and viscous simulations. (b) Total mass calculated by other codes
in inviscid simulations. The legend details were presented in de Val-Borro
(2006). (c) Total mass calculated by other codes in viscous simulations. See de
Val-Borro (2006) again for legend details. \label{figure 21}}
\end{figure}

Compared to the codes adopt absorbing or open inner boundary, our code
maintains relative high level of the total mass in the disk during the
simulation. After $200$ orbits evolution, the total mass in our simulation
decreases only $\sim2.5\%$ for the inviscid case and $\sim3.5\%$ for the
viscous case. While most results of the other codes are at the range of
$2\%-9\%$ for the inviscid simulations and $5\%-8\%$ for the viscous
simulations(see \textbf{Figure} \ref{figure 21}).

\begin{figure}
\epsscale{0.6} \plotone{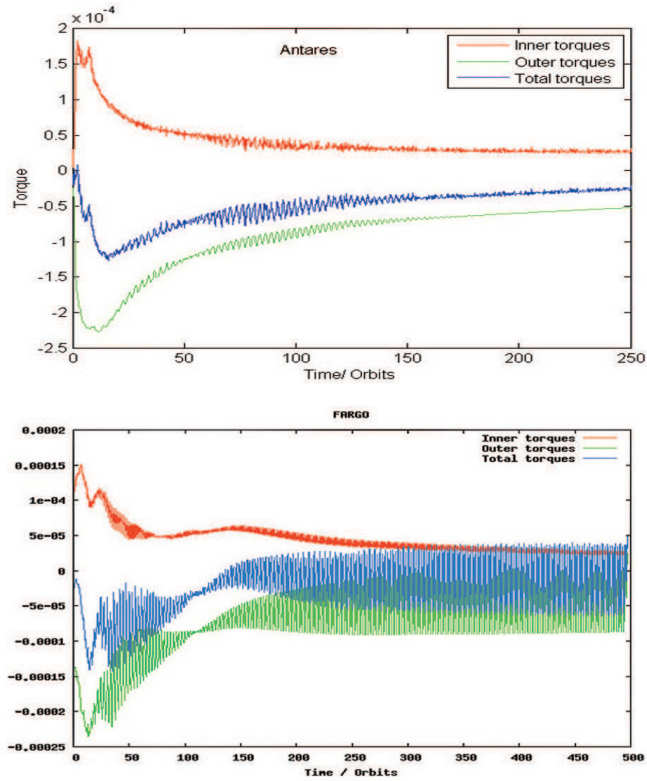} \caption{Comparison of the
time-averaged torques exerted on Jupiter in inviscid simulations. The results
of FARGO are obtained from the web:
http://www.astro.su.se/groups/planets/comparison, which is maintained by de
Val-Borro. \label{figure 22}}
\end{figure}

\begin{figure}
\epsscale{0.6} \plotone{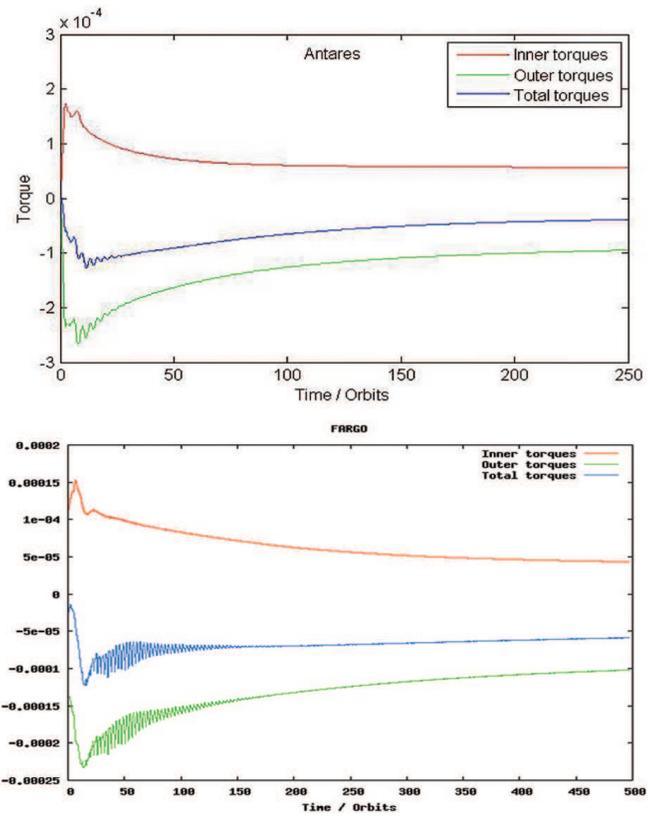} \caption{Comparison of the
time-averaged torques exerted on Jupiter in viscous simulations. The viscosity
coefficient is $\nu=10^{-5}$. The results of FARGO are obtained from the web:
http://www.astro.su.se/groups/planets/comparison, which is maintained by de
Val-Borro. \label{figure 23}}
\end{figure}

The comparisons of torques are showed in \textbf{Figure} \ref{figure 22}
(inviscid) and \textbf{Figure} \ref{figure 23} (viscous). Our results are
coherent well with FARGO both qualitatively and quantitatively. The averaged
total torque between $175-200$ is $-2.5\times10^{-5}$ in the inviscid
simulation and $-6.6\times10^{-5}$ in the viscous simulation. Both of this two
values are around the average of the results presented by \citet{Val06}.

According to the above comparisons, we conclude that our
code---\emph{Antares}---is reliable for simulating planet-disc interaction in
Cartesian grids. By adopting the CTU method and high resolution mesh, the
numerical dissipation (or grid effects) is negligible in our simulations.

\subsection{Initial Conditions}
The surface density of the gas disk ($\sigma$) varies as a function of its
radius $r$. As shown in \textbf{Table} \ref{table 2}, we adopt several
different initial distributions: $\sigma_0$, $\sigma_0 e^{-r^2/53}$, $\sigma_0
r^{-1/2}$ and so on. $\sigma_0$ is set to be 0.0006 in our units, which
corresponds to a height-integrated surface density $\sim 200 {\rm ~g/cm}^2$. The
angular velocity of the gas $u_\theta=r \Omega_g$ is slightly different from
the Keplerian velocity since the flow is in a centrifugal balance with both the
softened gravity of the star and the gas pressure which raises from the
distribution of the surface density $\sigma(r)$. The initial radial velocity of
gas is set to be 0. These initial conditions of the disk do not take into
account of the gravitational perturbation by the planets.

To set up a dynamical equilibrium in which the orbit of a planet is circular
and the stream lines are closed, we adopt an negligible initial mass for each
of the planets ($3\times10^{-7}$ or equivalently 0.1$M_\oplus$). And at the
very first 200 fixed circular orbits, both growth rates of the planets are
specified to be $\sim3.5\%$ per orbit until they achieve the mass of Jupiter
and Saturn, respectively. With this ``quiet-start" prescription, the planets
gain their masses through adiabatic growth so that the disk has enough time to
make a smooth response. The releasing consequence of the two planets is a
complicate issue, since it directly relates to the formation consequence of a
multiple planet system. We choose to release two planets at the same time, so
the pre-formed planet should stay at an circular orbit and wait for the later
one. Although this process may introduce some inconsistences, it's the most
suitable initial state to investigate the migration of a planet pair.

\section{Results}
\subsection{Effects of disk's surface density}
We consider a configuration in which Saturn locates outside the orbit of
Jupiter, which is similar to our Solar system. As shown in \citet{Mas01}, this
configuration usually leads to the convergent migration and resonance trapping
in a gas disk. In this paper we focus on the effects of the disk's surface
density. The slope of the surface density determines the magnitudes of torques
exerted on the planets, and thus affects the speed of the convergent migration
as well as the type and stability of MMRs. We assume the surface density of
disk is a function of its radii only: $\sigma=\sigma_0 r^{-\alpha}$ and obtain
a series of density distributions by varying the value of $\alpha$. As
summarized by \textbf{Table} \ref{table 2}, most of the typical $\alpha$ had
been considered: for example, a flat disk $\sigma=\sigma_0$, a moderate steep
disk $\sigma=\sigma_0 r^{-1}$ and some extreme steep ones, e.g.
$\sigma=\sigma_0 r^{-5/3}$.

\begin{figure}
\epsscale{0.6} \plotone{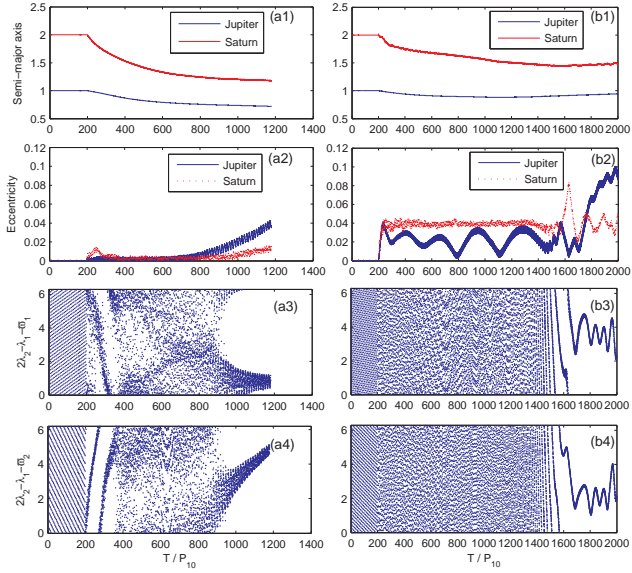} \caption{Orbital evolutions of
Jupiter and Saturn embedded in a flat disk where $\sigma=\sigma_0 r^{0}$
(Panels a1-a4) and a nearly flat disk where $\sigma=\sigma_0 e^{-r^2/53}$
(Panels b1-b4), respectively. (a1) and (b1) Evolutions of the semi-major axes
of the planets. The two planets approach to each other gently in a flat disk or
a nearly flat disk. The common migration stops or even reverses after the
establishment of resonance. (a2) and (b2) Evolutions of the eccentricities of
planets. (a3) and (a4) Evolution of the resonance
angles:$\theta_1=2\lambda_2-\lambda_1-\varpi_1$ and
$\theta_2=2\lambda_2-\lambda_1-\varpi_2$ of $2:1$ MMR. (b3) and (b4) Evolution
of the resonance angles: $\theta_1=2\lambda_2-\lambda_1-\varpi_1$ and
$\theta_2=2\lambda_2-\lambda_1-\varpi_2$ of $2:1$ MMR. (A color version of this
figure is available in the online journal.) \label{figure 3}}
\end{figure}

We start with a flat disk, where the initial surface density is
$\sigma_0=0.0006$ which corresponds to a height-integrated surface density
$\sim 200 {\rm ~g/cm}^2$. Panels (a1-a4) in \textbf{Figure} \ref{figure 3} show
the evolutions of the semi-major axes, eccentricities and resonant angles of
the two giant planets. When we release them at $t=200 P_{10}$, Jupiter had
almost opened a gap and begins a slow type II migration after a short
transition period. While the situation is quite different for Saturn: the tidal
torque generated by Jupiter keeps pushing the gas into the coorbital zone of
Saturn, thus greatly delays the gap opening process of Saturn. As a result,
Saturn migrates inward under the Lindblad and corotation torques, at a speed
much faster than that of Jupiter. After the convergent migration, Saturn is
trapped into the $2:1$ MMR with Jupiter. The eccentricities of both planets are
excited as soon as the MMR is established.

Following the flat disk, we run a moderate case in which the surface density of
disk is set to be $\sigma \sim e^{-\frac{r^2}{53}}$. This density profile is
derived from the analysis of \citet{Gui06} for an evolving disk. It is in fact
very close to the flat disk in our computational region($r\in(0,4)$). Embedding
in this kind of disk, the migration of Saturn is a little more oscillatory than
that in a flat disk, and we found the two planets stop and slightly reverse
their migration to outward after they had been locked into the $2:1$ MMR. These
are in good agreement with the results of \citet{Mor07} at the similar
parameter settings(the same disk aspect ratio, viscosity and surface density
profile), as shown in the Panels (b1-b4) of \textbf{Figure} \ref{figure 3}.

When the disk has a steep density profile, the situation becomes more complex.
According to our simulations, different values of $\alpha$ lead to different
rates of convergent migration, thus results in the trap of different MMRs. The
common migration speed and direction also depend on the slope of disk density.
We will state them in details as follows.

\begin{figure}
\epsscale{0.6} \plotone{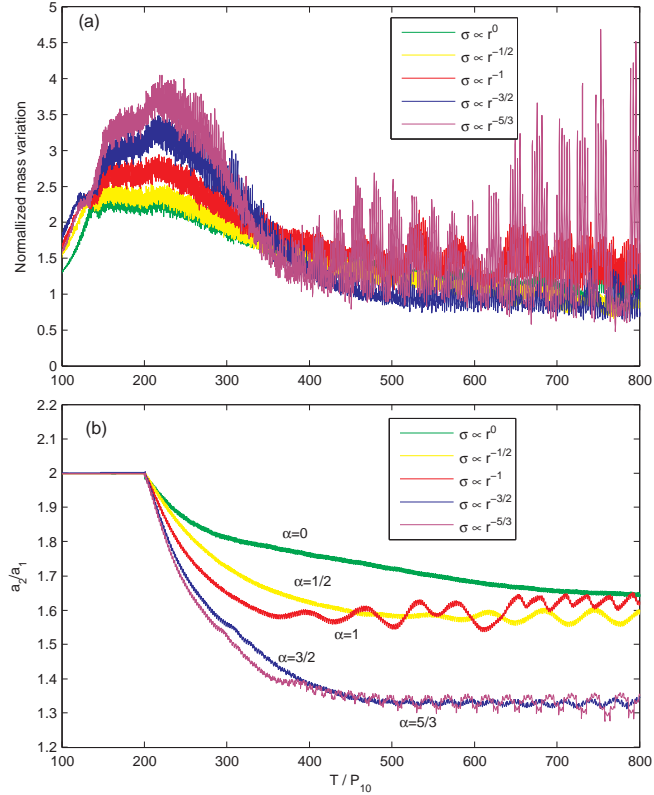}
\caption{ (a) Mass variation in the coorbital zone of Saturn during the
convergent migration stage. It decreases when $\alpha$ decreases ($\sigma
\propto r^{-\alpha}$)---from the top to the bottom in this panel, respectively.
The radii of coorbital region is set to be $1.5R_{Hill}$ and the mass variation
is normalized by the initial mass contains in Saturn's coorbital region. (b)
The different convergent migration rates with various $\alpha$: it increases at
large $\alpha$. The convergent migration is halted when the two planets are
locked into MMRs($2:1$ for $\alpha\leq1$ and $3:2$ for $\alpha>1$). Note that
mass variation decreases as the common gap forms. In Panel (a), the large
variations in the curve of $\sigma \propto r^{-5/3}$ when $T>400P_{10}$ are the
results of the increasing eccentricity of Saturn. This corresponds to the
oscillations of the convergent migration($\alpha=5/3$) in Panel (b). (A color
version of this figure is available in the online journal.) \label{figure 4}}
\end{figure}

First, the rate of convergent migration before the trapping of MMR increases as
the density slope $\alpha$ increases, as shown in Panel (b) of \textbf{Figure}
\ref{figure 4}. After the moment of release ($t=200P_{10}$), the
coorbital zone of Saturn is perturbed heavily by the density waves
generated by Jupiter. The migration of Saturn is thus dominated by the
corotation torque, which scales with the gradient of the potential vorticity
within the vicinity of the planet in the linear approximation\citep{Gol79,War91,War92}:
\begin{equation}
\Gamma_{C}\propto \sigma\frac{d\log{(\sigma/B})}{d\log{r}},
\end{equation}
where $B=\kappa^2/(4\Omega)$ is the second Oort constant with $B\sim r^{-3/2}$
in a nearly Keplerian disk, and $\sigma=\sigma_0 r^{-\alpha}$ is the surface
density. So the direction and magnitude of the corotation torque,
$\Gamma_{C}\propto (\frac{3}{2}-\alpha) r^{-\alpha}$, depends on $\alpha$. In
fact, when the planet is forced to migrate fast (e.g., scattered by other
planets) or its coorbital zone is perturbed heavily (in our cases, for example,
the density waves generated by Jupiter), the density gradient in its coorbital
zone is not monotone with $r$ and usually very complex. So the corotation
torque exerted on it should be evaluated by the real angular momentum exchanged
within the coorbital zone. \citet{Mas03} found a relation between the dramatic
migration rate and the coorbital mass deficit $\delta m$. In a Keplerian disk
it reads:
\begin{equation}
\frac{1}{2}a\Omega (M_p-\delta m)\dot{a}= \Delta\Gamma_{LR}-\frac{\pi a^2\delta m}{3x_s}\ddot{a},
\end{equation}
where $\Omega$ is the angular velocity of the planetary Keplerian motion, $M_p$
is the planet mass, $\Delta\Gamma_{LR}$ is the differential Lindblad torque and
$x_s$ is the half-width of planet coorbital zone. A runaway migration occurs
when $\delta m$ becomes comparable to the mass of planet $M_p$, because large
$\delta m$ leads to fast migration which breaks the assumption that $a$ stays
constant and results in an instability\citep{Mas03}. In our simulations, the
steeper density profile between the orbits of Jupiter and Saturn results in
heavier density waves that affect the coorbital zone of Saturn, and therefore,
generates greater $\delta m$. Panel (a) in \textbf{Figure} \ref{figure 4} shows
that the mass variation in the coorbital zone of Saturn increases at larger
$\alpha$.

On one hand, Saturn migrates inward faster at large $\alpha$. On the other
hand, according to the viscous evolution of disk, Jupiter migrates outward when
$\alpha>1/2$ (details are presented after \textbf{Equation} \ref{rvis}). So
Saturn and Jupiter migrate to each other faster when the surface density
becomes steeper, see Panel (b) in \textbf{Figure} \ref{figure 4}. As a result,
the time period that needed before the trapping of MMR between the two planets
is longer in flatter disk. For example, it will take 1000-1500 $P_{10}$ in a
disk with $\alpha <1/2$ (\textbf{Figure} \ref{figure 3}), while it needs only
300-500 $P_{10}$ in a disk with  $\alpha >1/2$(\textbf{Figure} \ref{figure 5}).

Second, The two planets may be trapped into different MMRs when the surface
density slope $\alpha$ varies. The phenomenon that different types of MMRs that
can be trapped during convergent migration depends on the migration speed is
revealed in \citet{Pap05}. In our cases, when the convergent migration is
relatively slow, i.e., $\alpha \leq 1$, Saturn is trapped by the $(p+1):p=2:1$
MMR of Jupiter(\textbf{Figure} \ref{figure 3},\ref{figure 5}). While
\textbf{Figure} \ref{figure 6} shows the results with a steeper disk where
$\alpha = 3/2$. The Saturn passes through the $2:1$ MMR and is trapped by the
$3:2$ MMR of the Jupiter. Large $p$ would be achieved by increasing the slope
of disk profile ($\alpha$), so that the speed of convergent migration is
increased. However, we do not observe any resonances with $p>2$ in our
simulations---$p=2$ is the highest value in our results. In fact, a transition
from $3:2$ MMR to $2:1$ MMR is observed when the surface density of disk is
very steep($\alpha = 5/3$), see \textbf{Figure} \ref{figure 7}. In this
simulation, Saturn first passes through the position of $2:1$ MMR with Jupiter
under fast inward migration and is then locked into the $3:2$ MMR with Jupiter.
When the resonance established, the two planets migrate outward together, and
at the mean time the migration of Saturn becomes unstable as its eccentricity
keeps growing. After several hundred orbits, when the eccentricity of Saturn
grows to $e > 0.15$, the $3:2$ MMR breaks and Saturn is scattered outward. As
soon as the resonance breaks, eccentricities of both the two planets are damped
effectively by the gas disk and then Saturn is captured by the $2:1$ MMR of
Jupiter. This result is consistent with that of \citet{Pie08}, and we find the
MMR is not so stable for high eccentricities of planets when they are embedded
in a steep disk($\alpha > 1$).

Third, the common migration speed and direction of the planet pair in MMR
varies with $\alpha$. The speed of common inward migration of the planet pair
slows down as $\alpha$ increases. When $\alpha > 1$, the two planets reverse
their migration to outward, and the migration speed increases as $\alpha$
increases, see \textbf{Figure} \ref{figure 8}. This phenomenon can be
understood as follows. In the case of one planet, Jupiter undergo type II
migration in the viscous disk. Each unit gas ring suffers a viscous torque
\begin{equation}
\Gamma_\nu=2\pi r^2 \nu \sigma  \frac{r d\Omega}{dr},
\end{equation}
where $\nu$ is the viscosity, $\sigma=\sigma_0 r^{-\alpha}$ is the surface
density of gas disk, $\Omega\sim r^{-3/2}$ is the angular velocity of
Kerplerian motion. Assuming a constant $\nu$ across the disk, the viscous
torque $\Gamma_\nu = -3\pi\nu\sigma_0 r^{1/2-\alpha}$. By further assuming the
disk remains Kelperian flow and integrating the motion equation in the
azimuthal direciton, the angular momentum transportation per unit mass is
governed by following formula:
\begin{equation}
\dot{r} \frac{d(r^2\Omega)}{dr}=\frac{1}{ 2\pi
r\sigma}\frac{\partial\Gamma_\nu}{\partial r}+\Lambda,
\end{equation}
where $\Lambda$ denotes the local injection rate of angular momentum per unit
mass into the disk gas from the planet\citep{Lin86b}. The effect of $\Lambda$
vanishes when the planet is treated as a part of the disk in type II migration.
Then we can obtain the radial movement of the gas under the effect of viscous
torque:
\begin{equation}
\dot{r}=\frac{1}{2\pi r \sigma}\frac{\partial\Gamma_\nu/\partial
r}{d(r^2\Omega)/dr}= -3\nu(\frac{1}{2}-\alpha)r^{-1}. \label{rvis}
\end{equation}
Since Jupiter now follows the movement of the gas, this in fact results in an
inward(or outward) migration of the planet at $\alpha<1/2$ ( $\alpha>1/2$,
respectively). It can be seen from Panels (a1) and (b1) in \textbf{Figure}
\ref{figure 5}, the initial stage ($t=200-500P_{10}$) of Jupiter evolution
(before it meets Saturn).

The presence of Saturn complicates the situation. Since Jupiter is massive than
Saturn, Saturn is in fact forced to migrate with Jupiter when
$\alpha>4/3$(\textbf{Figure} \ref{figure 6} and \ref{figure 7}). In the case of
$1/2 < \alpha < 1$, our simulation shows that the presence of Saturn perturbs
the previous outward migration of Jupiter, and makes it undergo slight inward
migration(see \textbf{Figure} \ref{figure 5}). The reason is that when Jupiter
encounters Saturn which is under fast inward migration, its outward migration
is slowed down. Then, the eccentricity of Jupiter is excited by the $2:1$ MMR.
As the orbit becomes eccentric, Jupiter cuts into the inner part of the gas
disk and this causes a negative corotation torque. \textbf{Figure} \ref{figure
9} shows the evolution of the torques exert on Jupiter and the mass variation
within the coorbital zone of it. It clearly shows that, after $t=700P_{10}$,
the large mass variation results in a negative corotation torque that reverses
the migration of Jupiter. The moment $t=700P_{10}$, is just the moment that the
eccentricity of Jupiter exceeds $0.17$ which is the critical value required by
Jupiter to cut into the inner gas disk (obtained by set $ae > 2.5R_{Hill}$,
with $R_{Hill}$ the Hill radius of Jupiter).

In this section we show that, for Jupiter and Saturn embedded in a gaseous
disk, their convergent migration rate, type of MMRs and subsequent common
migration depend on the density slope $\alpha$ of the gas disk. The direct
reason for these effects is the different torques(direction, value and type)
exert on the planets, which vary with $\alpha$.

\subsection{Torque analysis}
To understand the phenomena shown in previous section more deeply, we present
here some torque analysis based on the linear estimation. After the release of
planet at $t=200P_{10}$, the migration of Jupiter couples with the response of
gas disk. Since the coorbital zone of Jupiter is effectively cleared, the
angular momentum transportation between the inner and outer disk are mainly
done by the Lindblad torques. Due to the lack of Lindblad torques expression
for a disk under the perturbation of Jupiter, we estimate the differential
torque from linear estimation. The $m$th-order Lindblad torque can be expressed
as\citep{War97,Pap06}:
\begin{equation}
\Gamma^{LR}_m=\frac{{\rm sign}(\Omega_p-\Omega)\pi^2\sigma(r)}{3\Omega_p\Omega\sqrt{1+\xi^2}(1+4\xi^2)}\Psi^2,
\label{equation 2}
\end{equation}
where $\Omega,\Omega_p$ are the angular velocity at the specific location of
Lindblad resonance and the planets, respectively. $\xi$ is a function that
ensures the cutoff of the Lindblad torques at large $m$. This is naturally
satisfied when the density vanishes in the gap around the planet. $\Psi$ is the
force function which reads:
\begin{equation}
\Psi=r\frac{d\psi_m}{dr}+\frac{2m^2(\Omega-\Omega_p)}{\Omega}\psi_m. \label{Psi}
\end{equation}
By a high order interpolation and averaging along azimuthal direction, we
obtain the numerical density $\sigma(r)$ and angular velocity $\Omega(r)$
distribution in the disk at certain time. Then we determine the resonance
positions for each $m \leq 80$ as well as the Lindblad torque rises there
through \textbf{Equations} (\ref{equation 2}) and (\ref{Psi}).

In a flat disk, we find that the result follows the analytic prediction that
the outer Lindblad torque is always stronger, see \textbf{Figure} \ref{figure
10}. However, the existence of Saturn weakens the outer Lindblad torques
exerted on Jupiter by pushing gas outward away. And the inner disk is
strengthened by the steep density profile when $\alpha > 0$. \textbf{Figure}
\ref{figure 11} shows the differential Lindblad torque exerted on Jupiter at
the beginning of simulation ($t=0$), the moment of release ($t=200P_{10}$) and
the moment when the common gap has well formed ($1000P_{10}$), in a disk where
$\alpha=3/2$.  At the beginning of simulation, when the disk is still
unperturbed, the numerical results are fit well with the analytic results
except at $25 \le m\le 40 $. With the time passing by, both of the inner and
outer torques decrease and the position of maximum moves toward smaller $m$,
which corresponds to the gap formation process. At the release moment, when the
Saturn has already formed, the outer torque decreases more than the inner one
does, thus the net torque changes to positive. This positive torque drives
Jupiter to migrate outward. When the gaps of the two planets overlapped, the
inner and outer torques of Jupiter become comparable and are much smaller than
the initial value by an order of $2-3$. Then, Jupiter is in type II migration
effectively.

However, the migration of Saturn is dominated by the corotation torque. As we
have mentioned before, the coorbital zone of Saturn is always perturbed by
Jupiter when they are approaching to each other---even after the common gap has
formed(\textbf{Figure} \ref{figure 12}). The situation is more serious in a
steeper disk, because the density profile may amplify the density waves
generated by Jupiter. \textbf{Figure} \ref{figure 13} shows the torque
evolutions of both Jupiter and Saturn embedded in a gas disk where $\sigma \sim
r^{-3/2}$. Although the amplitude of corotation torque is comparable to that of
Lindblad torque exerted on the Jupiter, it oscillates around zero and its
average effect vanishes. The outer Lindblad torque is weakened by Saturn and
the net differential torque is slightly above zero.

At the mean time, the corotation torque exerted on Saturn overwhelms the
Lindblad torques. At the moment of release when $T=200P_{10}$, the corotation
torque is negative since the density gradient is not perturbed much and still
maintain negative around Saturn. Then, as the two planets approaching to each
other, the density gradient within the vicinity of Saturn changes to positive
and the corotation torque evolves to positive as well. At this moment, Saturn
lies close to the outer edge of the common gap, see \textbf{Figure} \ref{figure
14}. When $3:2$ MMR is established, the orbit of Saturn becomes more and more
eccentric. This makes Saturn cut through the outer edge of the gap
periodically, and generates periodical torques. When the eccentricity of Saturn
reaches $0.15$, it is scattered outward by the increasing corotation torque and
the resonance breaks(\textbf{Figure} \ref{figure 6} and \ref{figure 7}).

As shown by \textbf{Figure} \ref{figure 15}, the mass variation within the
coorbital zone of Saturn and the corotation torques exerts on it are strongly
correlated. In the plot, the short period corresponds to the orbit motion of
Saturn, as Saturn cuts into the outer disk in every orbit. And the long period
is the libration time of Saturn which reads:
\begin{equation}
t_{\rm lib}= \frac{2\pi a}{R_{co}}\left|\frac{1}{2}r\frac{\partial \Omega}{\partial r}\right|^{-1},
\end{equation}
where $R_{co}$ is the half width of the coorbital zone and usually equals to
$2.5$ times of the Hill radius of Saturn. The libration period at the position
of Saturn is $t_{\rm lib}\approx 32 P_{10}$, as shown in \textbf{Figure}
\ref{figure 15}, from  $T=1000P_{10}$ to $1100P_{10}$. This is also the period
that the coorbital mass exchanges its angular momentum with the
planet\citep{War91,Mas03}. The mass variations are normalized by the
unperturbed coorbital mass of each planet respectively. For Saturn, the peak
value immediately before the onset of instability is about 5 times greater than
its unperturbed state, which is roughly $10^{-4}$ in our unit, or $30\%$ of the
mass of Saturn. The angular momentum exchange between this part of coorbital
mass and Saturn results in the corotation torque that dominates Saturn's
migration.

In this section we show that the different migration rates and directions of
the two giant planets are the results of the combination of Lindblad and
corotation torques, which depend on the density slope of the disk ($\alpha$).
The corotation torque exerted on Saturn weakens the stability of Saturn's
orbit. The unstable migration is more serious in the disk with large $\alpha$,
see \textbf{Figure} \ref{figure 5}-\ref{figure 7}. And furthermore, the breaks
of MMR are observed when $\alpha \geq 3/2$(\textbf{Figure} \ref{figure 6} and
\ref{figure 7}). To analysis these breaks is helpful to understand the orbital
evolution following the convergent migration of Jupiter and Saturn. The details
and analysis are presented in the next section.

\subsection{Stability of MMRs}
MMRs are common results of the convergent migration of Jupiter and Saturn. In
this section, we investigate the stability of them. As showed in \textbf{Table}
\ref{table 2}, Saturn may be trapped by the $2:1$ MMR of Jupiter when the disk
is nearly flat and may reach the $3:2$ MMR if the surface density profile is
steep, e.g. $\alpha>1$. Furthermore, the $3:2$ MMR of the Jupiter and Saturn is
not stable as their eccentricities keep growing, while the $2:1$ MMR seems to
be more robust even at relatively high eccentricities $\sim 0.2$(e.g.,
\textbf{Figures} \ref{figure 3} and \ref{figure 5}).

The instability of planets in the $2:1$ or $3:2$ MMRs could be caused by the
overlap of two nearby resonances when the eccentricity is large enough. In the
case that both two planets are with equal masses ($m$) and move in circular
orbits, the overlap of nearby resonances occurs when the difference of their
semi-major axes is smaller than a limit\citep{Wis80,Gla93},
\begin{equation}
\frac{\Delta a}{a}\sim\frac{2}{3p} < 2(\frac{m}{M_*})^{2/7}, \label{chaos}
\end{equation}
where the relationship of $n_1/n_2=(p+1)/p=[(a+\Delta a)/a]^{3/2}\approx 1+3/2
(\Delta a /a)$ is used, with $n_1,n_2$ the mean motion of the two planets.
Assuming $m/M_*=(0.27-0.95)\times 10^{-3}$ for Jupiter and Saturn, the
application of \textbf{Equation} (\ref{chaos}) give $p>p_{\rm min}\sim 3-4$ for
the overlap between $(p+1):p$ and $(p+2):(p+1)$ MMRs.

However, when $p\le p_{min}$ their MMRs can also overlap if they are in high
eccentric orbits. In the framework of restricted three-body problem, a
planetesimal embedded in a gas disk will be trapped into the outer resonance of
a massive body. \citet*{KAR93} estimate of the minimum separation between MMRs
below which the planetesimals may undergo chaotic instability due to the MMR
overlap:
\begin{equation}
\frac{\Delta a}{a} \sim[(\frac{8 \pi m_p}{3 M_*})^2(\frac{3 v_{\rm dif}}{2 v_{\rm kep}})]^{1/9},
 \label{eq1}
\end{equation}
where $m_p$ refers to the mass of planet and $e$ is the orbital eccentricity of
planetesimal, $v_{\rm dif}$ is the difference between the gas velocity and the
local circular Keplerian velocity $v_{\rm kep}$. This velocity difference is a
result of the pressure rose by the slope of the surface density $\alpha$ in the
disk, which makes the gas circle the central star at a sub-Keplerian velocity,
and is referred as the drag effect of the gas disk. In a nearly flat disk,
$v_{\rm dif}/v_{\rm kep}$ is usually tiny, for example, $v_{\rm dif}/v_{\rm
kep}=0.002$ at $r=1$ when $\sigma \sim e^{-r^2/53}$, while it increases to
$\sim 0.01$ when $\alpha>3/2$. Use the approximation that $ 3/2 (\Delta
a/a)\approx 1/p \sim 1/(p+1)$ and the relation obtained by \citet{Wei85}:
\begin{equation}
e\approx (\frac{v_{\rm dif}/v_{\rm kep}}{p+1})^{1/2},
\end{equation}
we can get the minimum eccentricity, above which the $(p+1):p$ and
$(p+2):(p+1)$  MMRs will cross, from \textbf{Equation} (\ref{eq1}):
\begin{equation}
e_{min} \sim \frac{3 M_*}{8 \pi m_p}(\frac{2}{3}\frac{1}{p+1})^5. \label{emin}
\end{equation}
This gives $e_{min} \approx 0.065$ for $p+1:p=3:2$ MMR, and $\approx 0.49$ for
2:1 for a Jupiter mass disturber. Although \textbf{Equation} (\ref{emin}) is
obtained from the restrictive three-body problem, our simulation show that it
also applies to Jupiter-Saturn case. In fact, we find the $3:2$ MMR breaks as
soon as the eccentricity of Saturn reaches $0.15$ and larger $\alpha$ doesn't
change this value but only makes instability happen earlier, see
\textbf{Figure} \ref{figure 6} and \ref{figure 7}. The $2:1$ MMR is much more
robust, because it only requires $e\leq 0.5$ to maintain stable and the
eccentricity excited by $2:1$ always stay at $e\leq0.2$, see \textbf{Figures}
\ref{figure 3} and \ref{figure 5}.

The other reason for the instability of MMR is the runaway migration induced by
the corotation torque. We had shown that, a moderate planet will undergo
runaway migration(Type III migration) when its coorbital zone is perturbed
heavily by the density waves generated by the other giant
planet\citep{Zha08,Zha08b}. It is believed that two giant planets would be
trapped and migrate together when their gaps overlapped well. It's true that
when the density distribution of disk is nearly flat, the convergent migration
is relatively slow and Saturn has enough time to clean its coorbital zone
before it interacts with Jupiter directly. However, when the disk is steep, say
$\sigma \sim r^{-3/2}$, the convergent migration is so fast that the Jupiter
would catch Saturn to the $3:2$ MMR before the later one opens a clean gap, see
\textbf{Figure} \ref{figure 16}. At this time, Saturn is forced to migrate with
Jupiter while the corotation torque still dominates its migration. As we can
see from \textbf{Figure} \ref{figure 6}, the migration curve becomes very
oscillatory as soon as Saturn is trapped by the MMR with Jupiter.

In fact, eccentricity plays an essential role on triggering runaway migration
even if a clean gap formed around the planet orbit. As we mentioned in previous
section, as soon as the giant planet undergoes eccentric motion, it cuts
through the edge of the gap periodically when $a e\geq R_{\rm gap}$, where
$R_{\rm gap}$ is the width of the gap. Then, a periodic torque rises as a
result of the replenishment of gas into the vicinity of the planet. This torque
may change the direction of planet's migration (see \textbf{Figure} \ref{figure
9}) or kick the planet away quickly. A good approximation for the gap width is
$R_{\rm gap}\approx2.5R_{\rm Hill}$, the same with the radius of the coorbital
zone of a giant planet by assuming the viscosity is low, where $R_{\rm Hill}$
is the Hill radius of this planet. In the case of Saturn, it requires $e \sim
0.15$ to allow Saturn cut into the disk. However, as Saturn lies outside the
orbit of Jupiter, it can only touch the outer edge of the common gap, where the
surface density gradient is positive and the corotation torque exerted on it is
positive as well, see \textbf{Figure} \ref{figure 14}. Thus, Saturn tends to be
scattered outward at this kind of configuration.

According to our results and analysis, we find that the instability of MMR is
mainly due to the large eccentricity excited by the MMR itself. High
eccentricity either leads to the overlap of MMRs or makes the planet cut into
the disk and generates strong corotation torque. As the eccentricity evolutions
of the two planets locked in resonance vary with the type of
resonance\citep{Mic06}, density slope ($\alpha$) will affect the final orbit
configuration of the planet pair.

\section{Conclusions and discussions}

A series of high resolution hydrodynamic simulations have been performed to
investigate the orbital evolution of Jupiter and Saturn embedded in a
protostellar disk. We focus on the effects of different surface density
profiles of the gas disk where $\sigma \propto r^{-\alpha}$. The instability of
mean motion resonance caused by the convergent migration is also studied.
According to the results and analysis of our simulation, we summarize our
conclusions as follows, as well as the discussions and implications for these
results.

(1) The existence of Jupiter(massive planet) delays the gap formation process
of Saturn(light planet) and generates great perturbations within the coorbital
zone of Saturn. These perturbations result in the inward or outward runaway
migration(type III migration) of Saturn, depending on $\alpha$, the density
slope of the gas disk.

The effects of the pre-formed giants are very important for understanding the
orbital evolution of a multiple planet system. To investigate these effects, we
fix the very first $200P_{10}$ orbits of both the two planets when they are
growing from $0.1$ Earth mass to 1 Jupiter and 1 Saturn mass respectively, then
we release them at the same time. This, in fact, assumes that the Jupiter had
formed before the Saturn did since the gap of Jupiter had well formed while the
Saturn's hadn't yet.

The orbital evolutions of the planets form later are greatly affected by the
pre-formed giant ones. The first generation of giant planets strongly modified
their neighborhood by opening gaps at several Hill radius from themselves
\citep{Bry00}, as well as the planetesimal gaps \citep{Zho07}. \citet*{Pie08}
had shown that the light planets will be trapped at the edge of the giant
planet's gap. This halt is because of the balance between the Lindblad torque
and the corotation torque which rises at the density jump. This happens when
the mass of the outer planet is low: $m_2 \leq 20M_\oplus$. For massive outer
planets,i.e., $m_2\sim M_J$, the strong tidal effect guarantees the gap
formation process and slow type II migration could be expected. The situation
becomes complex when the outer planet has a moderate mass, $m_2\sim M_S$ which
is critical to open a gap. Its orbital evolution thus strongly depends on the
initial conditions, for example, the initial density profile of the disk
$\alpha$.

\citet*{Zha08} had shown that the existence of the gaseous disk expands the
interaction region of the proto-planets. As we have shown in the previous
section, the tidal effect of Jupiter keeps pushing the gas away from its
coorbital zone and replenish the gap of Saturn (see \textbf{Figure} \ref{figure
12} and \ref{figure 14}). This effect increases as $\alpha$
increases(\textbf{Figure} \ref{figure 4}) and drives the orbital evolution of
the system unstable. If the initial separation is large or the disk is nearly
flat, clear gaps would form around the planets' orbit and gentle convergent
migration could be expected. This will result in a less compact and more stable
system.

(2) The convergent migration rate is proportional to the surface density slope
$\alpha$. And, as a result, the types of MMRs depend on $\alpha$ as well. The
two planets approach to each other gently when $\alpha < 1$, and are locked
into $2:1$ MMR. When $\alpha > 1$, the convergent migration is fast and the
$3:2$ MMR is reached.

The convergent migration rate is one of the most critical issues to determine
the consequential migration of the planet pair. Many factors may affect this
rate, for example, the masses of the two planets \citep{Pie08}, the viscosity
and the disk aspect ratio \citep{Mor07}. However, the essential factor which
results in differential migration is the various torques by which the planets
are driven. These torques intensively relate to the surface density of the disk
in which the planets embedded. For a low mass planet, the differential Lindblad
torque is negative and the value is proportional to the density slope
$\alpha$($\sigma\sim r^{-\alpha}$)\citep{War91}. The migration of massive
planet is determined by the global distribution of the angular momentum(relates
to $\alpha$) and the viscosity of the gaseous disk. The corotation torque
relates to the vortensity gradient(relates to $\alpha$ as well) within the
coorbital zone of planet\citep{Mas03}. The analysis of torques by a
semi-analytical method have been shown in the previous section, see
\textbf{Figure} \ref{figure 9}-\ref{figure 11} and \ref{figure 13}.

To get the different convergent rate naturally, we choose a series of density
profiles of the disk where $\sigma\sim r^{-\alpha}$. Most of the typical
density profiles have been adopted, see \textbf{Table} \ref{table 2}. Our
results show that the convergent migration rate increases as $\alpha$
increases, see \textbf{Figure} \ref{figure 4}. This is reasonable since the
migration of Saturn is accelerated by the steep density slope while the
migration of Jupiter turns outward when $\alpha>1/2$. The type of resonance is
determined by how close the two planets could approach. When the disk is nearly
flat, where $\alpha<1$, the $2:1$ MMR is a robust outcome. While the $3:2$ MMR
is more favorite in the steep disk where $\alpha>1$. The disk with a density
profile around $\alpha \approx 1$ shows a transition phenomena and the high
eccentricity of Jupiter reverses its outward migration to inward, see
\textbf{Figure} \ref{figure 5} and \ref{figure 9}.

(3) The $3:2$ MMR of the two planets is unstable when the eccentricity of
Saturn becomes large enough in a steep disk where $\alpha>3/2$. We estimate
that the critical value is $e_s\sim0.15$ with our settings.

The $3:2$ MMR of Jupiter and Saturn is thought to be robust for many settings,
but we find that this configuration may breaks down when the eccentricity of Saturn
grows too high in a gaseous disk where the surface density gradient is pretty
steep $\alpha>3/2$. In fact, the eccentricities will be exited as soon as the
resonance established no matter the disk is nearly flat or very steep
(\textbf{Figures} \ref{figure 3} and \ref{figure 5}-\ref{figure 7}). However, in a
steep disk, the situations are different. First, the convergent migration is
much faster. This enables the two giant planets get much closer and as a
result, the dynamical instability becomes possible, for example, the overlap
of resonances. Second, caused by the fast convergent migration as well, the
establishment of resonance occurs before a clear common gap formed. This will
result in a strong corotation torque which may drive an unstable migration of Saturn,
 see \textbf{Figure} \ref{figure 6}. Third, the steep gradient of
density in fact amplifies the density waves propagating from Jupiter to Saturn
and results in relatively heavier perturbation in coorbital zone of Saturn.

In our simulations,  the planet pair migrates outward when $\alpha>1$. The separation
between them increases as they preserve the $3:2$ MMR. Thus, Saturn is pushed
outward further and further. At the mean time, the growing eccentricity makes
Saturn cut into the outer disk deeper and deeper. The density gradient at the
out edge of the common gap is positive and generates positive corotation torque
that pushes Saturn outward further. When the eccentricity is high enough,
Saturn is scattered outward. Our results show that the critical value is
$e\geq0.15\sim0.2$ for $3:2$ MMR of Jupiter and Saturn.

We also find that the onset of instability would be suspend when $\alpha$
decreases. In fact, in the case where $\alpha=4/3$, the two planets maintain
$3:2$ MMR over $2000P_{10}$, see \textbf{Figure} \ref{figure 8}. So the long
time stability could also be expected by choosing a proper $\alpha$, and this
still needs further simulations. The $2:1$ MMR seems to be more stable for high
eccentricities and we find the two planets could be re-locked into $2:1$ MMR
just after the break of $3:2$ MMR when they are both migrating outward, see
\textbf{Figure} \ref{figure 7}.

If the initial density profile of our Solar nebular is relatively steep, e.g.
$\alpha > 4/3$, then the formation of the main configuration of our Solar
system could be this way: (a) Jupiter and Saturn first migrate inward when
their masses are still low. (b) By accreting gas from the nearby disk, they
grow up quickly. While Jupiter grows much faster than Saturn does through the
runaway accretion, and the mass difference results in a convergent migration
between them. (c) Then the $3:2$ MMR of Jupiter and Saturn is established and
their migration turns outward with the resonance preserved. (d) When the
eccentricity of Saturn becomes too high, the resonance breaks down and Saturn
is scattered outward to the place near its present location or re-captured by
the $2:1$ MMR of Jupiter. Neptune and Uranus are also scattered out away at the
mean time. (e) The eccentricities of both the inner rocky planets and the outer
gas giant are damped effectively by the gas disk after the instability. (f)
After the dissipation of gas, the planets evolve to their present locations by
the interaction with the planetesimal disk. Of course, many details are need to
be addressed especially for the orbital evolutions of the inner rocky planets.
However, compared to the results of \citet{Mor07b}, our results suggest that
the instability would happens before the dissipation of gas. The remaining gas
may corresponds to the low eccentricities of the main planets in our Solar
system. And since Jupiter migrates inward before its outward migration, the
final location is not far from its birth place.

We also notice that the excitation of Jupiter's eccentricity is previous than
that of Saturn in $2:1$ MMR and is laggard in $3:2$ MMR. Since the eccentricity
is a critical issue to guarantee the stability of Jupiter-Saturn pair, its
evolution and constraints need to be addressed in details by considering the
effect of the interacting disk. The results of a reverse orbital configuration
of Saturn and Jupiter is in preparing as well.

\section{Acknowledgement}
We thank D.N.C. Lin, A.Crida, W. Kley for their constructive conversations. We
also thank M. de Val-Borro for providing data for the comparison. Zhou is very
grateful for the hospitality of Issac Newton institute during the Program
`Dynamics of Discs and Planets'. This work is supported by NSFC (Nos.
10925313,10833001,10778603), National Basic Research Program of
China(2007CB814800) and Research Fund for the Doctoral Program of Higher
Education of China (20090091120025).

{}
\clearpage

\begin{deluxetable}{lcccccr}
\tablecaption{A short list of extra-solar planets in resonance.\label{table 1}} \tablehead{ \colhead{System} &
\colhead{ No.} & \colhead{$P$ [day]} & \colhead{$M\sin i$ [$M_J$]} & \colhead{$a$ [$AU$]} & \colhead{$e$} &
\colhead{Refs.}} \startdata
GJ 876  & c & 30.57 & 0.56 & 0.13 & 0.2 & 1 \\
(2:1)   & b & 60.13 & 1.89 & 0.21 & 0.04 & \\
\tableline
HD 128311& b & 458 & 2.3 & 1.08 & 0.23 &  2\\
(2:1)    & c & 918 & 3.1 & 1.71 & 0.22 & \\
\tableline
HD 73526& b & 188 & 2.90 & 0.66 & 0.19 & 3 \\
(2:1)   & c & 377 & 2.50 & 1.05 & 0.14 & \\
\tableline
HD 82943& b & 217.9 & 1.4 & 0.74 & 0.46  & 4 \\
(2:1)   & c & 456.6 & 1.78& 1.19 & 0.36 & \\
\tableline
HD 160691& d & 310.55 & 0.52 & 0.92 & 0.067 & 5\\
(2:1)    & b & 643.25 & 1.68 & 1.5  & 0.13 & \\
\tableline
HD 45364& b & 227 & 0.19 & 0.15 & 0.17 & 6\\
(3:2)   & c & 343 & 0.66 & 0.68 & 0.09 & \\
\tableline
HD 60532& b & 201 & 3.15 & 0.76 & 0.27 & 7\\
(3:1)   & c & 605 & 7.46 & 1.58 & 0.038 & \\
\tableline
\enddata
\tablecomments{{\bf References.} (1) Marcy et al. 2001;(2) Vogt et al. 2005; (3) Tinney et al. 2006; (4) Lee et
al. 2006; (5) Gozdziewski et al. 2007; (6) Correia et al. 2009; (7) Laskar \& Correia 2009.}
\end{deluxetable}

\begin{deluxetable}{cccccc}
\tablecaption{A summary of our simulations.\label{table 2}} \tablehead{ \colhead{Case} & \colhead{Configuration}
& \colhead{Direction } & \colhead{$\sigma$} & \colhead{Resonance} & \colhead{Instability}} \startdata
1  & J-S & inward & $\sigma \sim r^0$ & $2:1$ & no \\
\tableline
2  & J-S & inward & $\sigma \sim e^{-\frac{r^2}{53}}$ & $2:1$ & no \\
\tableline
3  & J-S & inward & $\sigma \sim r^{-\frac{1}{2}}$ & $2:1$ & no \\
\tableline
4  & J-S& inward & $\sigma \sim r^{-\frac{2}{3}}$ & $2:1$ & no \\
\tableline
5  & J-S & inward & $\sigma \sim r^{-1}$ & $2:1$ & no \\
\tableline
6  & J-S & outward & $\sigma \sim r^{-\frac{4}{3}}$ & $3:2$ & no \\
\tableline
7  & J-S & outward & $\sigma \sim r^{-\frac{3}{2}}$ & $3:2$ & yes \\
\tableline
8  & J-S & outward & $\sigma \sim r^{-\frac{5}{3}}$ & $3:2$ $\rightarrow$ $2:1$ & yes \\
\tableline
\enddata
\end{deluxetable}

\clearpage

\begin{figure}
\epsscale{0.6} \plotone{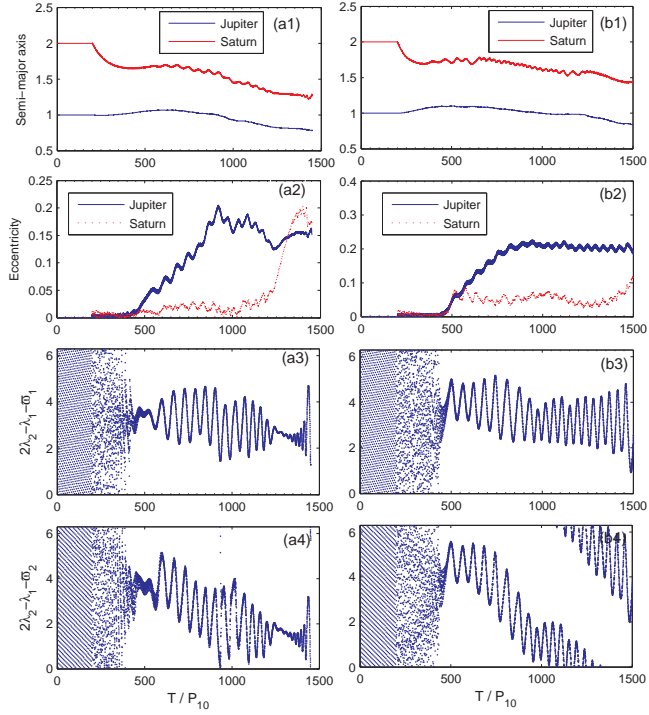} \caption{Orbital evolutions
of Jupiter and Saturn embedded in a slightly steep disk where $\sigma=\sigma_0
r^{-1/2}$ (Panels a1-a4) and a steeper disk where $\sigma=\sigma_0 r^{-1}$
(Panels b1-b4).  (a1) and (b1) Evolution of the semi-major axes of the two
planets. Jupiter first migrates outward when $\alpha>1/2$ ($\sigma\propto
r^{-\alpha}$) and then reverses its migration to inward after locking into 2:1
MMR with Saturn.  (a2) and (b2) Evolutions of the eccentricities of the
planets. Note that the Jupiter's eccentricity is excited heavily by the
resonance. (a3) and (a4) Evolution of the resonance angles:
$\theta_1=2\lambda_2-\lambda_1-\varpi_1$ and
$\theta_2=2\lambda_2-\lambda_1-\varpi_2$ of $2:1$ MMR. (b3) and (b4) Evolution
of the resonance angles: $\theta_1=2\lambda_2-\lambda_1-\varpi_1$ and
$\theta_2=2\lambda_2-\lambda_1-\varpi_2$ of $2:1$ MMR. Due to the effects of
the dissipative disk, the libration center changed from
$(\theta_1,\theta_2)=(0^\circ,0^\circ)\rightarrow$ asymmetric libration.(A
color version of this figure is available in the online journal.) \label{figure
5}}
\end{figure}

\begin{figure}
\epsscale{0.6} \plotone{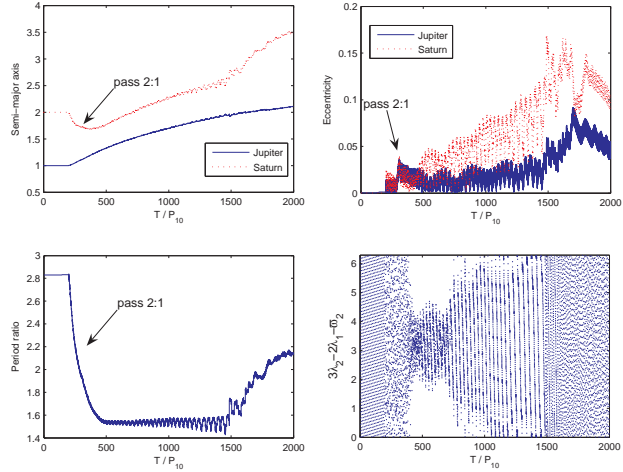} \caption{Orbital
evolutions of Jupiter and Saturn embedded in a disk where $\sigma=\sigma_0
r^{-3/2}$. Saturn passes through the position of $2:1$ MMR of Jupiter and then
is catched by the $3:2$ MMR. Scattering happens at $T=1500P_{10}$ when
$e_s\geq0.15$. (a) Evolutions of the semi-major axes of the two planets.
Migration of Saturn becomes unstable after it is trapped into resonance with
Jupiter. (b) Evolutions of the eccentricities of the planets. The excitation of
Jupiter's eccentricity by the $3:2$ MMR is behind that of Saturn. (c)
Evolutions of the period ratio $P_S/P_J=(a_2/a_1)^{3/2}$. (d) Evolutions of the
resonance angle $\theta\lambda_2-2\lambda_1-\varpi_2$ of $3:2$ MMR. The
resonance becomes unstable as the eccentricities keep growing. (A color version
of this figure is available in the online journal.) \label{figure 6}}
\end{figure}

\begin{figure}
\epsscale{0.6} \plotone{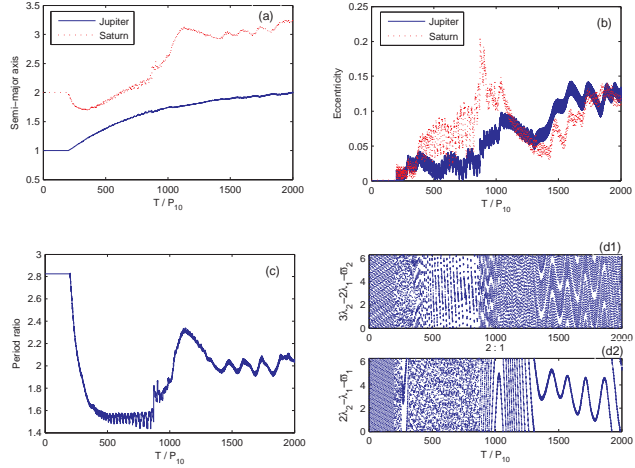} \caption{Orbital
evolutions of Jupiter and Saturn embedded in a disk where $\sigma=\sigma_0
r^{-5/3}$. Saturn passes through the position of $2:1$ MMR of Jupiter and then
is catched by the $3:2$ MMR. Scattering happens at $T=800P_{10}$ when
$e_s\geq0.15$. Eccentricities decrease rapidly after the break of resonance and
then the Saturn is trapped by the $2:1$ MMR of the Jupiter at $T=1400P_{10}$ .
(a) Evolutions of the semi-major axes of the two planets. (b) Evolutions of the
eccentricities of the planets. It clearly shows that the excitation of
Jupiter's eccentricity is previous than that of Saturn in $2:1$ MMR and is
laggard in $3:2$ MMR. (c) Evolutions of the period ratio
$P_S/P_J=(a_2/a_1)^{3/2}$. One can see the re-capture of $2:1$ MMR. (d1) and
(d2) Evolutions of the resonance angle: $\theta=3\lambda_2-2\lambda_1-\varpi_2$
of $3:2$ MMR(d1) and $\theta=2\lambda_2-\lambda_1-\varpi_1$ of $2:1$ MMR(d2).
(A color version of this figure is available in the online journal.)
\label{figure 7}}
\end{figure}

\begin{figure}
\epsscale{0.6} \plotone{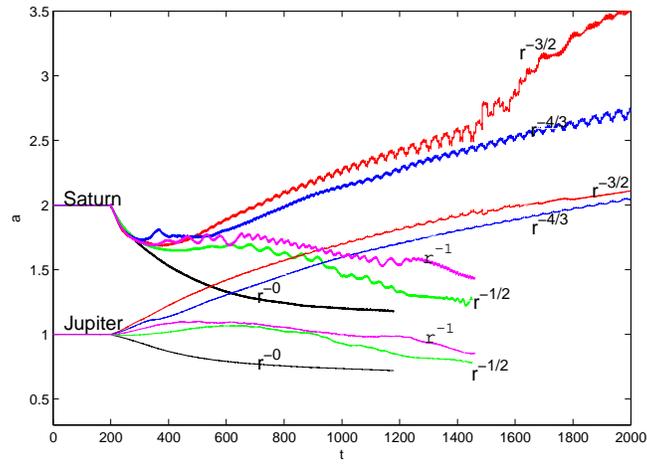} \caption{Migrations of Jupiter and
Saturn embedded in different disk where the density slope $\alpha$ varies from
$0$ to $3/2$. The two planets migrate inward when $\alpha<\frac{1}{2}$, and
migrate outward when $\alpha>1$. It shows a transitional state when
$\frac{1}{2}< \alpha <1$. (A color version of this figure is available in the
online journal.) \label{figure 8}}
\end{figure}

\begin{figure}
\epsscale{0.6} \plotone{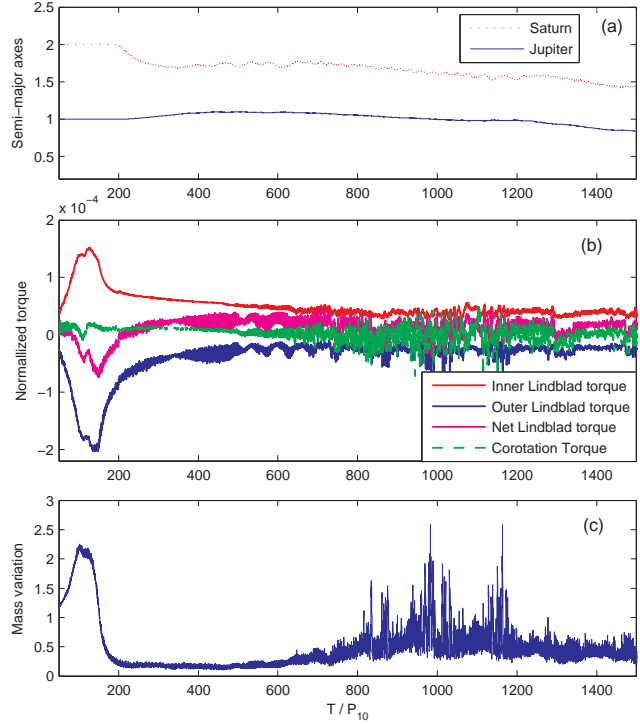} \caption{ (a)
Evolutions of the semi-major axes of Jupiter and Saturn. (b) Evolutions of the
inner, outer and net Lindblad torques as well as the corotation torque exerted
on Jupiter. The net Lindblad torque is positive at the release moment
($t=200P_{10}$), then a negative corotation torque rises and dominates
Jupiter's migration($t>700P_{10}$). (c) Mass variation within the coorbital
zone of Jupiter. The large oscillations after $t=700P_{10}$ result in the
negative corotation torque which reverse the migration of Jupiter. The surface
density profile is $\sigma=\sigma_0 r^{-1}$ in this simulation. (A color
version of this figure is available in the online journal.) \label{figure 9}}
\end{figure}

\clearpage
\begin{figure}
\epsscale{1} \plotone{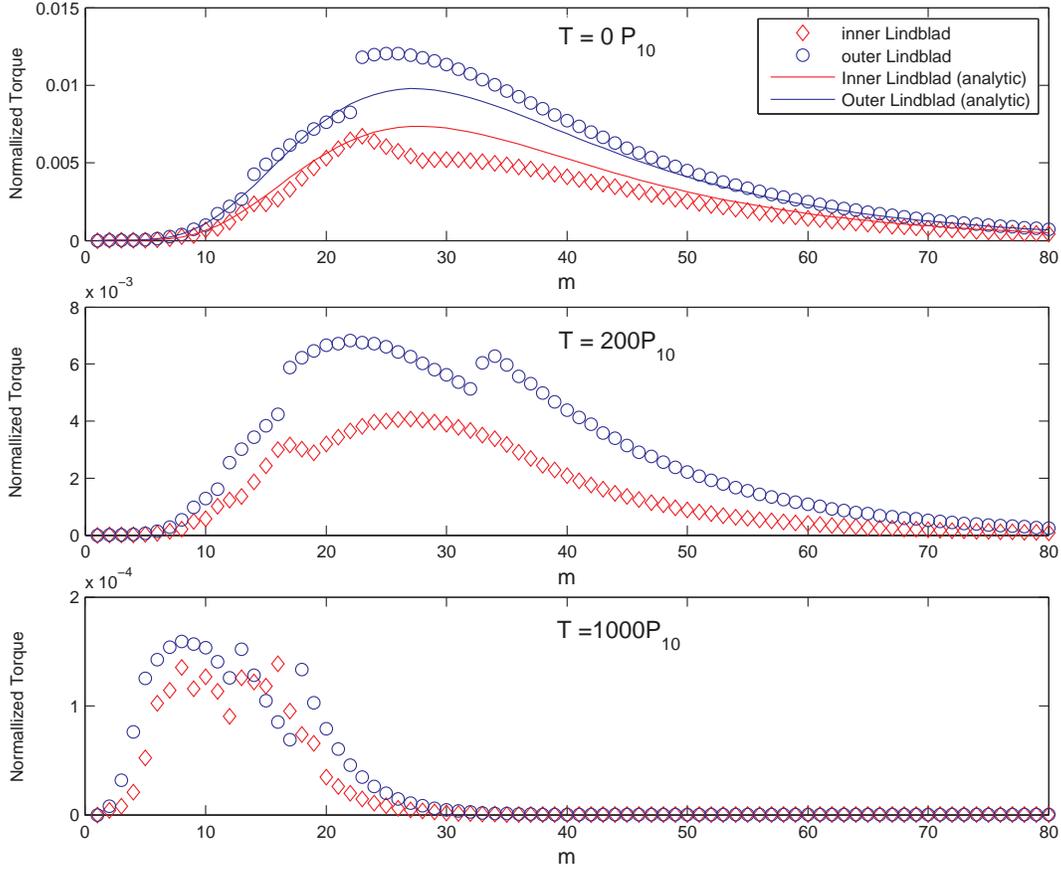} \caption{The $m$-th inner and
outer Lindblad torques exerted on Jupiter at different times, where $|m|=2-80$.
The density profile of the disk in which the two planets embedded is
$\sigma=\sigma_0 r^{0}$. From the top to the bottom, the time points are the
initial moment, the release moment and the moment after common gap has formed,
respectively. At initial moment, the torques are calculated as if Jupiter was
embedded in an unperturbed disk. The solid lines denote the analytic results
while the diamonds and circles denote the semi-analytic results. Note that the
outer Lindblad torque is always larger than the inner one and the position of
maximum value moves left as the gap becomes deeper and wider. The kinks in the
numerical results indicate the sudden changes of density or angular velocity
distribution of gas, or the failures of locating the exact positions of
resonances( extrapolation value is adopted instead), which is mainly due to the
lack of radial resolution of the numerical data.\label{figure 10}}
\end{figure}

\begin{figure}
\epsscale{1} \plotone{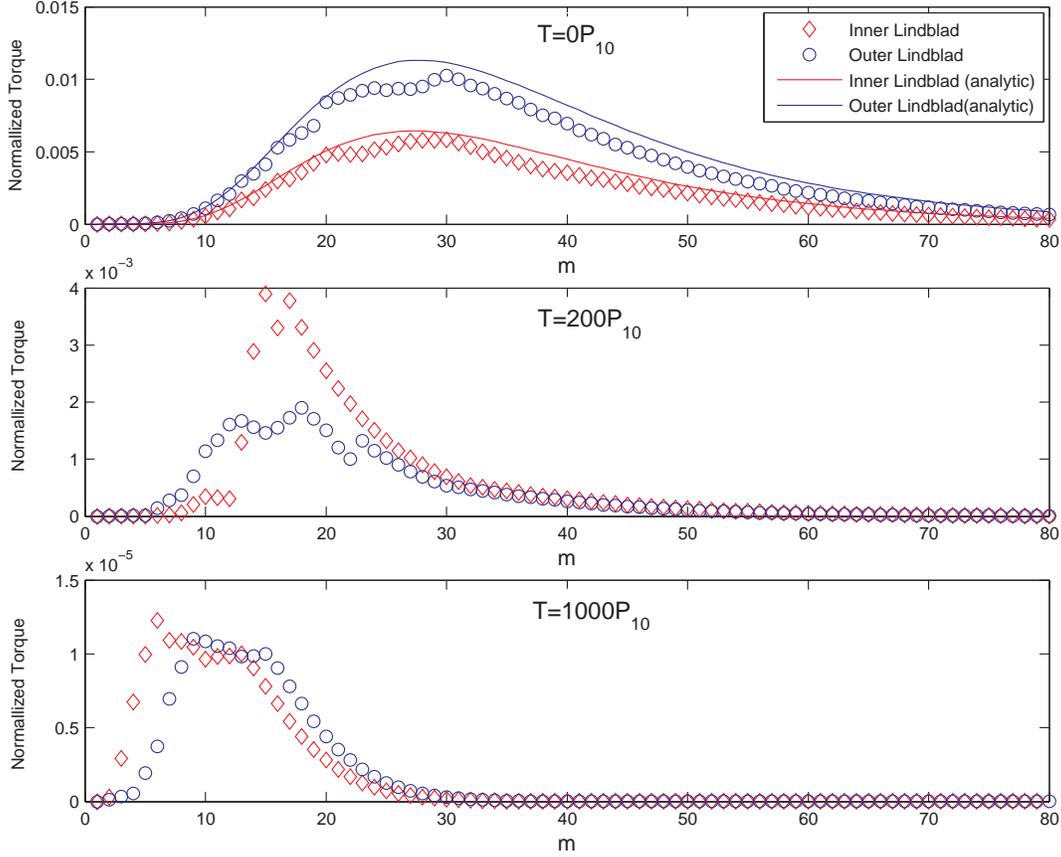} \caption{The $m$-th($|m|=2-80$)
inner and outer Lindblad torques exerted on Jupiter which embedded in a disk
where $\sigma=\sigma_0 r^{-3/2}$. From the top to the bottom, the time points
are the initial moment, the release moment and the moment after common gap has
formed, respectively. At initial moment, the torques are calculated as if
Jupiter was embedded in an unperturbed disk. The solid lines denote the
analytic results while the diamonds and circles denote the semi-analytic
results. Note that the inner Lindblad torque becomes larger than the outer one
at the release moment. Compared to \textbf{Figure} \ref{figure 10}, one may
find that the inner Lindblad torque almost remains the same level while the
outer one decreases a lot. This could be explained by the mass reduction in the
outer disk caused by Saturn. \label{figure 11}}
\end{figure}

\begin{figure}
\epsscale{0.6} \plotone{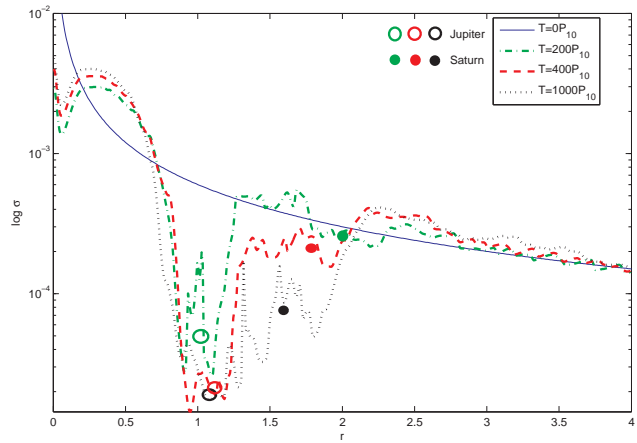} \caption{This figure
show the cross section of surface density at different moments. The initial
density profile is $\sigma=\sigma_0 r^{-1}$. One can see the overlapping
process of two gaps. Note that the gap formation of Saturn is much delayed($T
\approx1000P_{10}$)and the variation of density is acute at Saturn's vicinity.
The coordinates of planets in the figure denote their positions (x axis) and
the average gas density around them (y axis). (A color version of this figure
is available in the online journal.)\label{figure 12}}
\end{figure}

\begin{figure}
\epsscale{0.6} \plotone{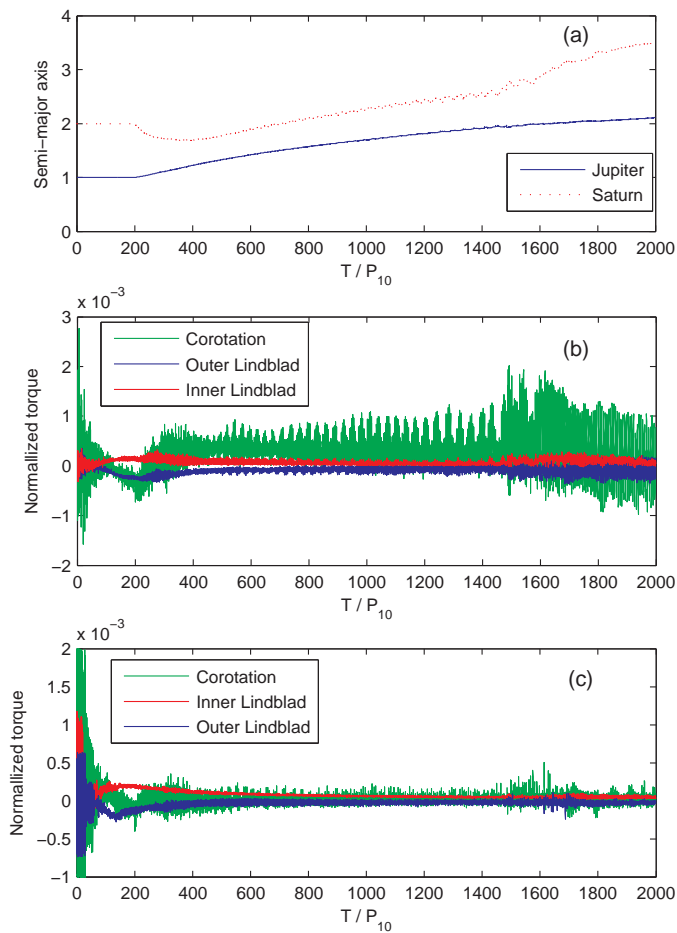} \caption{ (a)
Evolution of the semi-major axes of Jupiter and Saturn. (b) Evolution of the
inner/outer Lindblad torques and the corotation torque exerted on Saturn. It
shows clearly that the migration of Saturn is dominated by the corotation
torque. (c) Evolution of the inner/outer Lindblad torques and the corotation
torque exerted on Jupiter. The average value of corotation torque vanishes and
the migration of Jupiter is dominated by the Lindblad torques. The surface
density profile is $\sigma=\sigma_0 r^{-3/2}$. (A color version of this figure
is available in the online journal.)\label{figure 13}}
\end{figure}

\begin{figure}
\epsscale{0.6} \plotone{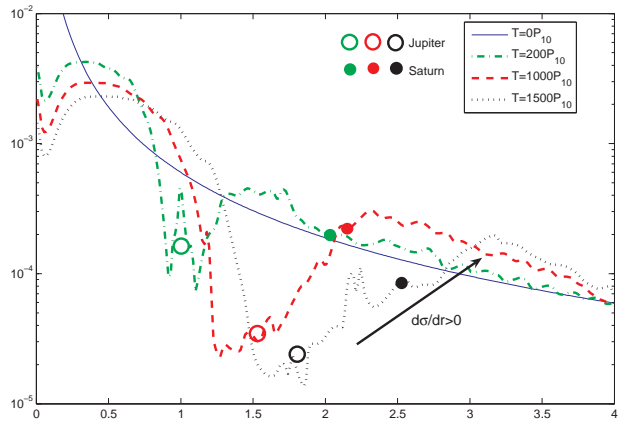} \caption{This figure
shows the cross section of surface density at different moments. The initial
density profile is $\sigma=\sigma_0 r^{-3/2}$. The common gap forms after
$1000P_{10}$ evolution time. One may notice that the density gradient is
positive $\frac{d \sigma}{d r}>0$ at the out edge of the common gap, which
produces positive corotation torque who drives Saturn outward away. The
coordinates of planets in the figure denote their positions ($x$ axis) and the
average gas density around them ($y$ axis). (A color version of this figure is
available in the online journal.)\label{figure 14}}
\end{figure}

\begin{figure}
\epsscale{0.6} \plotone{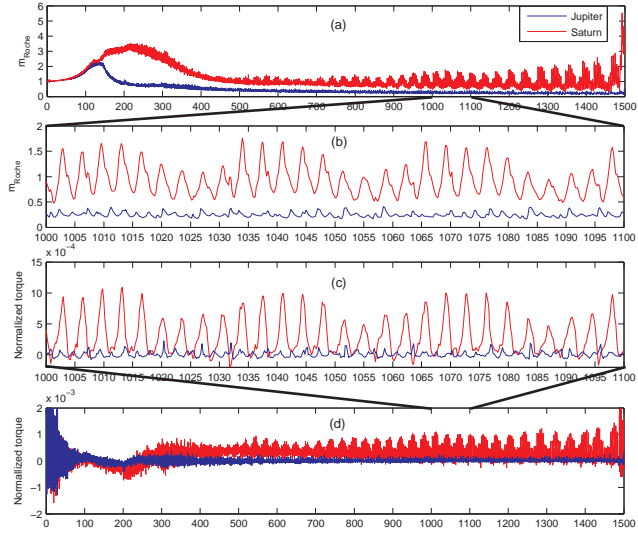} \caption{ Panel (a) shows
the evolution of mass variations within the coorbital zone of Jupiter and
Saturn. Panel (b) zooms in from $T=1000P_{10}$ to $T=1100P_{10}$. Panel (c)
shows the evolution of corotation torque. Panel (d) zooms in from
$T=1000P_{10}$ to $T=1100P_{10}$ too. The initial surface density profile is
$\sigma=\sigma_0 r^{-3/2}$. One may find the oscillations of the mass variation
and the corotation torque match well. The short period of oscillations
correspond to the orbit period and the long period is the libration period of
the horse-shoe orbit, which equals to $32P_{10}$ at the position of Saturn. (A
color version of this figure is available in the online journal.)\label{figure
15}}
\end{figure}

\begin{figure}
\epsscale{0.6} \plotone{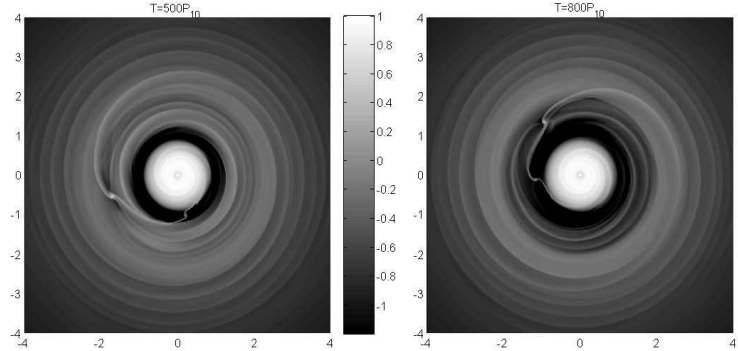} \caption{Density contours
at $T=500P_{10}$ and $T=800P_{10}$ in a disk where $\sigma=\sigma_0 r^{-3/2}$.
One may find the Jupiter has opened a clear gap while the Saturn is still
surrounded by gas at $T=500P_{10}$ (left figure). And the shock waves generated
by Jupiter perturbs the coorbital zone of Saturn when the two gaps are
overlapping $T=800P_{10}$ (right figure). \label{figure 16}}
\end{figure}
\end{document}